\documentclass[aps,pre,preprint,superscriptaddress,amsmath,amssymb]{revtex4-2}
\usepackage{graphicx}
\usepackage{dcolumn}
\usepackage{bm}
\usepackage[mathlines]{lineno}
\usepackage{epsfig}
\usepackage{xcolor,colortbl}
\usepackage{tabularx}
\usepackage{epstopdf}
\usepackage{amsmath,amssymb,amsthm}
\usepackage{float}
\usepackage{mathtools}
\usepackage{subfig}
\usepackage{hhline}
\usepackage{algcompatible}
\usepackage{newfloat}
\usepackage[justification=RaggedRight]{caption}
\usepackage[pdftex,bookmarks=true,colorlinks=false]{hyperref}
\DeclareFloatingEnvironment[
fileext=loa,
listname=List of Algorithms,
name=ALGORITHM,
placement=tbhp,
]{algorithm}

\usepackage{multirow}
\usepackage{geometry}
\usepackage{natbib}
\usepackage{enumerate}
\usepackage{enumitem}
\usepackage{titlesec}
\usepackage{wrapfig}
\usepackage{lipsum}
\usepackage{newtxtext}
\usepackage[varvw]{newtxmath}
\usepackage{indentfirst}
\usepackage{comment}
\graphicspath{{figures/}} 
\usepackage[nottoc]{tocbibind}
\usepackage{titleps}
\usepackage{dsfont}
\usepackage{relsize}
\usepackage[hang,flushmargin]{footmisc} 
\usepackage{listings}
\usepackage{setspace}
\usepackage[labelfont=bf]{caption}
\usepackage{amsfonts}
\usepackage{booktabs}
\usepackage{siunitx}
\usepackage{rotating}
\usepackage{cleveref}

\newcommand{\cepsilon}{\mathcal{E}} 
\newcommand{\dd}{\mathrm{d}} 
\newcommand{\dif}{\mathcal{D}}

\titleformat{\section}[hang]{\large\bfseries}{\arabic{section}{. }}{0pt}{\large\bfseries}
\titleformat{\subsection}[hang]{\normalsize\bfseries}{\arabic{section}.\arabic{subsection}{. }}{0pt}{\normalsize\bfseries}

\makeatletter
\g@addto@macro\normalsize{%
	\setlength\abovedisplayskip{4pt}%
	\setlength\belowdisplayskip{16pt}%
	\setlength\abovedisplayshortskip{1pt}%
	\setlength\belowdisplayshortskip{15pt}%
}

\begin{document}
\title{
Neural networks for turbulent transport prediction in a simplified model of tokamak plasmas}
\author{L. M. Pomârjanschi}
\email{ligia.pomarjanschi@inflpr.ro}
\affiliation{National Institute of Laser, Plasma and Radiation Physics,	M\u{a}gurele, Bucharest, Romania}
\affiliation{Faculty of Physics, University of Bucharest,	M\u{a}gurele, Bucharest, Romania}
\date{\today}

\begin{abstract}
	The method of using neural networks (NNs) for turbulent transport prediction in a simplified model of tokamak plasmas is explored. The NNs are trained on a database obtained via test-particle simulations of a transport model in the slab-geometrical approximation. It consists of a five-dimensional input of transport model parameters and the radial diffusion coefficient as output. The NNs display fast and efficient convergence, a validation error below 2$\%$, and predictions in excellent agreement with the real data, obtained orders of magnitude faster than test-particle simulations. In comparison to a spline interpolation, the NN outperforms, exhibiting better predicting and extrapolating capabilities. We demonstrate the preciseness and efficiency of this method as a proof-of-concept, establishing a promising approach for future, more comprehensive research on the use of NNs for transport predictions in tokamak plasmas.
\end{abstract}

\keywords{artificial neural network, fusion plasma, turbulent transport}

\maketitle

\section{Introduction}
\label{s0}

Nuclear fusion is one of the most promising solutions for a clean and reliable energy source to meet the world's current energy demands. Consequently, there is a significant and ongoing interest in developing a functional nuclear fusion reactor. For almost 70 years, the scientific community has struggled with the challenges of nuclear fusion, with confinement being the central issue.

Turbulence, while present in many physical systems such as the atmosphere, the oceans, and astrophysical media, gives rise to one of the main difficulties regarding the confinement of plasma within a nuclear reactor, namely, the turbulent transport. In such environments, turbulence is mainly represented by the turbulent electric potential, and it leads to one of the primary transport mechanisms of energy and particles. The large-scale, drift-type instabilities have the most significant impact on the dynamics of charged particles, particularly the ion-temperature-gradient (ITG) and the trapped-electron-mode (TEM).

The simplest way to characterize the transport of matter or heat within the paradigm of local transport is through the macroscopic transport coefficients, i.e. the diffusion and the average velocity. The regimes discussed in this work, while anomalous, are purely diffusive and do not exhibit sub- or super-diffusive behavior, i.e. non-local transport.

Considering its complexity and the inherently stochastic nature of turbulence, the approaches to studying turbulent transport are numerical in nature. The main method for studying plasma dynamics, turbulence evolution, and transport is through gyrokinetic simulations \cite{gyro_1,gyro_2,gyro_3}. This technique is used to describe the collective behavior of the plasma in the approximation of the particle \textit{gyromotion}, by solving an associated kinetic equation coupled to a Poisson-like equation. Although precise, gyrokinetic simulations are computationally demanding and require significant computational resources, in part due to the matter-fields self-consistency of the problem.

Another approach to the issue of turbulent transport is through test-particle simulations, or direct numerical simulations (DNS) \cite{palade_fast, palade_w}. The working principle is to follow individual particles in an ensemble of given electromagnetic configurations, not taking into consideration the collective interaction of the plasma with the electromagnetic field; using these trajectories, inferences on the transport coefficients can be made. They are significantly more time-efficient and less computationally demanding than gyrokinetic simulations due to the removal of self-consistency, and are convenient for studying the confinement of particles in various configurations and regimes. For these reasons, in the present work we use test-particle simulations to evaluate the turbulent transport in tokamak plasmas.

Nonetheless, the model for test-particle simulations depends on turbulence parameters, which are largely unknown, and plasma parameters, which must be varied in order to accommodate for different tokamak devices and turbulence states. Therefore, although this approach is faster than gyrokinetic simulations, it is far too slow for applications where the transport coefficients could be useful, such as integrated modeling and real-time control applications \cite{real_time_modeling_1}. One promising solution for this problem lies in the rapidly-evolving field of neural networks.

In the recent years, there have been substantial advancements in machine learning; neural networks (NNs) are a particular branch of this domain. NNs are collections of interconnected \textit{neurons} organized in \textit{layers}, which are able to ``learn" through extensive training and comprehensive processing of existing data, and, afterwards, make predictions in order to perform tasks such as pattern recognition, classification, or nonlinear regression. As a brief summary of the working principle of NNs: to each neuron-neuron connection there corresponds a \textit{weight}, and to each individual neuron, a \textit{bias}; the preexisting data fed to the NN is structured into ``inputs" and ``outputs", and the purpose of the trained NN is to predict the ``output", given the ``input"; during the training, the weights and biases of the NN are iteratively adjusted in order to replicate the real data as closely as possible.

\begin{figure}
	\centering
	\includegraphics[width=0.6\linewidth]{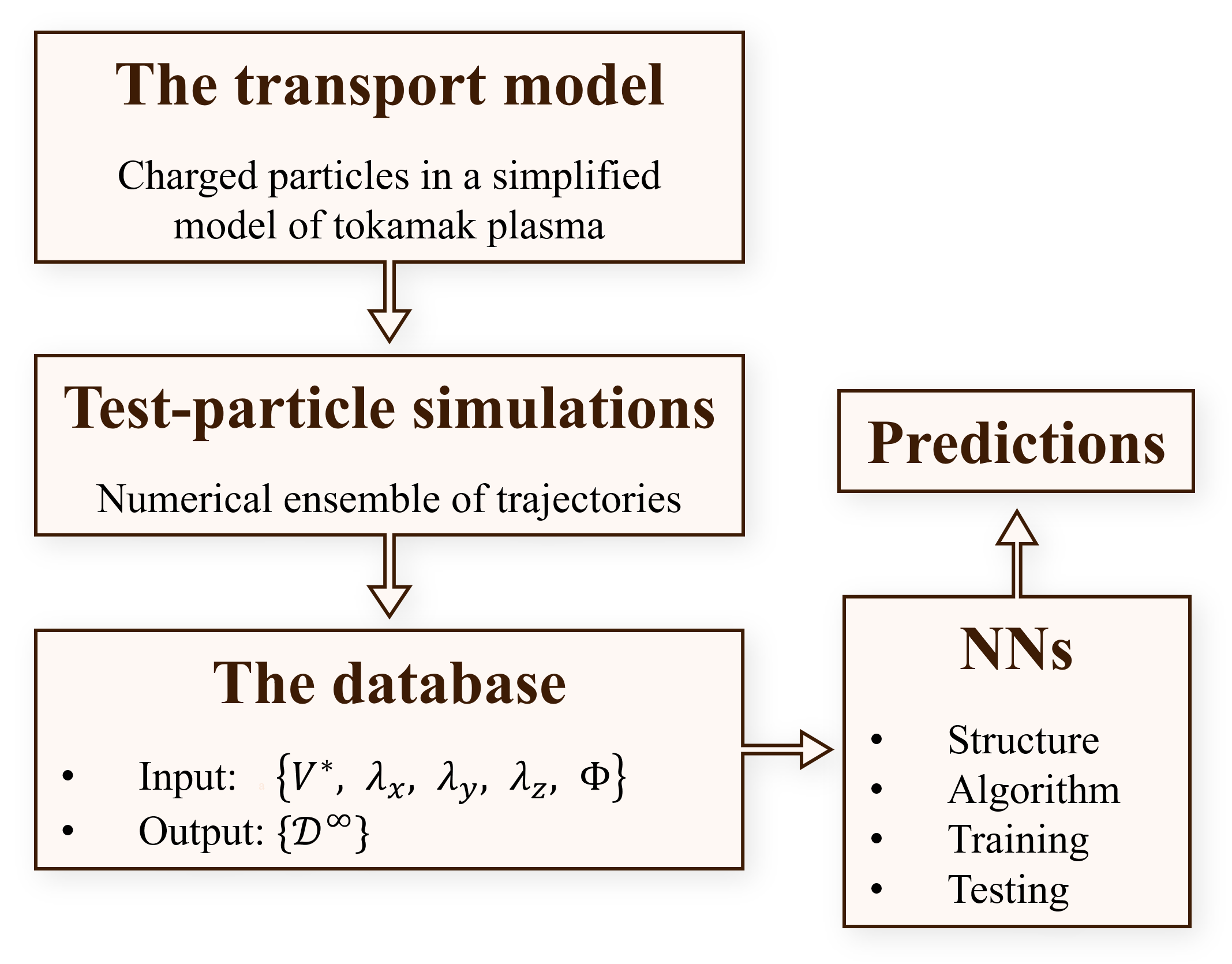}
	\caption{ \centering Schematic preview of the NN building components approached in this paper.}
	\label{fig_-1}
\end{figure}

In this work, we aim to use NNs as a tool for predicting turbulent transport in a simplified model of tokamak fusion plasmas. In order to train the NN, we construct a database with inputs consisting of plasma and tokamak parameters, and outputs consisting of asymptotic diffusion coefficients. A schematic preview of the NN building components approached in this paper is shown in Figure \ref{fig_-1}. It must be noted that the scope of this paper is a proof-of-principle, focusing solely on constructing the database and training the NN; it is meant to underline the precision and viability of future uses of this technique. Hence, we wish to lay the building blocks for future works in which we intend to use a more robust model of tokamak plasma dynamics, and construct a more versatile training database (i.e. which can accommodate additional transport regimes and more realistic tokamak devices).

The rest of the manuscript is structured as follows: in Section \ref{s1}, we describe the transport model (\ref{s1.1}), the numerical methods through which we study the system (\ref{s1.2}), and the NN setup (\ref{s1.3}); Section \ref{s2} outlines the numerical details; in Section \ref{s3}, we present the results, and Section \ref{s4} addresses the conclusions and outlook.

\section{Theory}
\label{s1}
\subsection{The transport model}
\label{s1.1}

We consider a tokamak plasma configuration in the slab-geometrical approximation; this allows us to simplify the equations of motion (EOMs) of charged particles, while capturing the most important details of the dynamics. The EOMs are described in field-aligned coordinates $(x,y,z)$ \cite{palade_peaking} related to the radial, poloidal and toroidal coordinates $\left( r,\theta,\varphi \right)$ as follows:

\begin{align}\label{}
	\begin{split}
		x&=r-r_0,\\
		y&=r_0\left( \theta -\frac{\varphi}{q(0)} \right),\\
		z&=R_0q(0)\theta,
	\end{split}
\end{align}

\noindent  with $q(0)$ -- the safety factor evaluated at $x=0$; $\;r_0 = a_0/2$; and $R_0,a_0$ -- the major and minor radii of the tokamak. $(x,y,z)$ are obtained in the large-aspect-ratio limit $\left( a_0/R_0\ll1 \right)$ of the natural field-aligned coordinate system. The plasma is immersed in a strong, constant magnetic field, which is oriented along the $z$-direction (also denoted as ``parallel"): $\textbf{B} = B_0\cdot \hat{e}_z$, with $\hat{e}_z$ -- the contravariant versor along $Oz$. In this slab-geometrical limit, we can write the EOMs for the dynamics of ions of mass $m$ and charge $q$, in the presence of a turbulent electric potential $\phi_g(\textbf{x},t)$, in the guiding-center approximation, with the particles' trajectories are described by $(\textbf{x}_\perp,x_\parallel, v_\parallel,\mu)$, as follows:

\begin{alignat}{2}
		\frac{d \textbf{x}_\perp}{d t} & = \frac{1}{B}\hat{e}_z\times\nabla\phi_g(\textbf{x},t)-\frac{m}{q B^2}\partial_t\nabla_\perp \phi_g(\textbf{x},t)-\frac{1}{qB}\left(m v_\parallel^2+\mu\right)\cdot\hat{e}_y, \label{1.1} \\
		\frac{d x_\parallel}{d t}&=v_\parallel , \label{1.2} \\
		\frac{d v_\parallel}{d t}&=-\frac{q}{m}\partial_z \phi_g(\mathbf{x},t) ,\label{1.3}\\
		\frac{\dd \mu}{\dd t}&=0.  \label{1.4}
\end{alignat}

The ``$\perp$" in equation \eqref{1.1} denotes the perpendicular, $(x,y)$ direction, and we recognize the first two terms as the $\textbf{E}\times\textbf{B}$ and the polarization drifts, respectively. The last term of the equation is a simplified version of the curvature and grad-B drifts, in the slab approximation, evaluated at the low-field side of the plasma, $\left\{x=0, \;y=0\right\}\equiv\left\{r=a/2 ,\;\theta=0 \right\}$, in the equatorial plane \cite{madi_hidden}, with $\mu=mv_\perp^2/2B$ -- the magnetic moment, and $v_\perp$ -- the particle velocity in the perpendicular plane.

In order to account for the effects of a finite Larmor radius, the turbulent potential of eqs. \eqref{1.1}-\eqref{1.4} is gyro-averaged, $\phi_g(\textbf{x},t)=\langle \phi(\textbf{x},t)\rangle_g$ \cite{palade_fast_ions, larmor_avg_1, larmor_avg_2,palade_alpha}. The turbulent electric potential is modeled as a zero-averaged, homogeneous, Gaussian random field, with the following Fourier representation \cite{palade_fast}:

\begin{equation}\label{2}
	\phi_g(\mathbf{x},t) =A_\Phi \int \hspace{-6pt} \dd \mathbf{k} \;S^{1/2}(\mathbf{k})\zeta(\mathbf{k}) e^{i(\mathbf{k}\cdot\mathbf{x}-\omega(\mathbf{k})t)}J_0 \left( k_\perp \rho \right),
\end{equation}

\noindent with $A_\Phi$ -- turbulence amplitude; $\rho=mv_\perp/(qB)$ -- Larmor radius; $\textbf{k}\equiv(k_x,k_y,k_z)$ -- wavenumbers, $k_\perp\equiv\sqrt{k_x^2+k_y^2}$; $\;J_0$ -- Bessel function of the first kind; $\omega(\textbf{k})$ -- frequencies; $S(\textbf{k})$ -- turbulence spectrum; and $\zeta(\textbf{k})$ -- Gaussian white noise, with $\langle\zeta(\textbf{k})\rangle=0$, $\langle\zeta(\textbf{k})\zeta(\textbf{k}')\rangle=\delta(\textbf{k}+\textbf{k}')$. Note that $J_0$ is a consequence of the gyro-averaging of the turbulent electric potential.

The turbulence spectrum $S(\textbf{k})$ represents the Fourier transform of the Eulerian autocorrelation function of the turbulent potential, $\cepsilon(\phi(\textbf{x},t),\phi(\textbf{x}',t'))=\langle \phi(\textbf{x},t)\phi(\textbf{x}',t') \rangle $.  To simplify the model, we only take into account an ITG-driven turbulence spectrum $S(\textbf{k})$ (considering the TEM components negligible). Experimental evidence and gyrokinetic simulations \cite{ITG_spectrum_exp_1, ITG_spectrum_exp_2, ITG_spectrum_exp_3, ITG_spectrum_exp_4} have shown that the ITG turbulence spectrum exhibits a single peaked structure along the radial and parallel directions, and a double-peaked structure along the $y$-direction; therefore, we use an analytical form of $S(\textbf{k})$ \cite{ITG_spectrum_analytical,palade_alpha} that is in accordance with these observations:

\begin{equation}\label{3}	
	S(\textbf{k}) \;\propto\; e^{-\frac{k_x^2\cdot\lambda_x^2}{2}-\frac{k_z^2\cdot\lambda_z^2}{2}} \; \frac{k_y}{k_0}\left(e^{-\frac{\left(k_y-k_0\right)^2\lambda_y^2}{2}}-e^{-\frac{\left(k_y+k_0\right)^2\lambda_y^2}{2}}\right).
\end{equation}

 The parameters $\lambda_x, \lambda_y$ and $ \lambda_z$ represent the correlation lengths along their respective directions, and $k_0$ has an influence on the positions of the two symmetrical maxima of the $k_y$ spectrum. We assume that the frequencies follow a linear dispersion relation, $\omega(\textbf{k}) \propto \textbf{k}\cdot\textbf{V}^{*}$, where $\textbf{V}^{*}$ is the ion diamagnetic drift velocity, driven by the pressure gradient, such that $\textbf{V}^{*} = -\mathbf{\nabla} p\times\textbf{B}/(n_i q B^2)$. By expressing it in terms of the ITG scale-length, $L_{T_i}^{-1}\equiv \lvert \nabla T_i/T_i \rvert$, we obtain $ \textbf{V}^* = -\hat{e}_y\cdot v_{th}\rho_i/L_{T_i} $, and the corresponding linear dispersion relation:
 
  \begin{equation}\label{4}
 	\omega(\textbf{k}) \approx  - \textbf{k}  \cdot \textbf{V}^*\;\frac{T_e}{T_i}\frac{1}{1+\left(T_e/T_i\right)^2\left(\rho_i k_\perp\right)^2}  \to \frac{- k_y V^* }{1+\left(\rho_i k_\perp\right)^2} ,
 \end{equation}
 
\noindent the latter of which was obtained by setting the electron temperature equal to the ion temperature, $T_e \approx T_i$.

We consider a single species of particles with mass number $A$ and charge number $Z$, in thermal and collisional ionization equilibrium with the bulk plasma. Consequently, the particles' kinetic energies, $\;E_k=\mu B+\frac{1}{2}m v_\parallel^2$,  follow a Boltzmann distribution, $f(E_k) \propto \sqrt{E_k}e^{-\frac{E_k}{T}}$, while the distribution of the initial pitch angles, $\xi \equiv \arctan (v_\perp/v_\parallel)$, is uniform on the interval $\left[-\pi,\pi\right]$. 

For numerical purposes, we scale the EOMs as follows: 
$\left(x,y,\lambda_x,\lambda_y\right)\to\rho_i$;\,
$\left(z,\lambda_z\right)\to a_0$;  			\,
$\left(k_x,k_y\right)\to \rho_i^{-1}$;			\,
$k_z\to a_0^{-1}$;\; $t\to a_0/v_{th}$;			\,
$v_\parallel \to v_{th}$; 						\,
$\phi \to A_\Phi$; 								\,
$V^*\to v_{th} \rho_i/a_0$; 					\,
$B\to B_0$;										\,
$m\to m_i$; 									\,
$q \to e$;										\,
$E_k\to T_i$,									\,
with 
$v_{th} = \sqrt{T_i/m_i}, \; \rho_i = m_i v_{th}/(\rvert e \lvert B_0)$ -- the thermal velocity and the Larmor radius of the ions; \,
$\Phi\equiv \lvert e \rvert A_\Phi/T_i$; 		\,
$B_0$ -- the magnetic field strength near-axis; 
and considering only the species of hydrogen ions, with mass $m_i$, charge $e$, and $A=Z=1$.

We are interested in the radial diffusion coefficient:
 
 \begin{align}
 \dif_{rr}(t)\equiv\dif(t)=\frac{1}{2}\frac{d}{d t} \langle \langle x(t)^2\rangle-\langle x(t)\rangle^2 \rangle. \label{1.0}
 \end{align}
 
For the scope of this paper, we limit ourselves to \textit{diffusive} regimes of transport, for which we compute the asymptotic radial diffusion coefficient, $\lim\limits_{t\to\infty}\dif(t)\equiv \dif^\infty$.

\subsection{The statistical approach and numerical implementation}
\label{s1.2}

In order to investigate the turbulent transport, we implement the transport model described in Section \ref{s1.1} using a test-particle method, or Direct Numerical Simulations (DNS) \cite{palade_fast, palade_parallel}. This exact-in-principle method mimics real trajectories $\textbf{x}(t)$ resulting from the EOMs \eqref{1.1}-\eqref{1.4} in different turbulent realizations; this is achieved by computing the trajectories using an explicit representation of the turbulent fields. 

The  turbulent electric potentials are constructed as an ensemble of dimension $N_p$ of stochastic, zero-averaged, homogeneous random fields $\left\{\phi(\textbf{x},t)\right\}$. The effects of intermittency on the distribution of the turbulent potential have already been studied in a previous work \cite{palade_pom_2022} and have been found to be minimal; thus, we can assume the Gaussianity of the fields. The ensemble of potentials $\left\{\phi(\textbf{x},t)\right\}$ drives an associated ensemble of trajectories $\left\{\textbf{x}(t)\right\}$ according to the EOMs \eqref{1.1}--\eqref{1.4}; the transport coefficients are then computed as Lagrangian statistical averages over the resulting trajectories. We use a discrete Fourier representation of the potential \cite{palade_fast}:

\begin{equation}\label{5}
	\phi_g(\mathbf{x},t) = \sqrt{\frac{2}{N_c}}\sum_{i=1}^{N_c} J_0(k_{\perp}^i \rho) \sin \left(\mathbf{k}_i \mathbf{x}- \omega(\mathbf{k}_i) t+\zeta(\textbf{k}_i) \right), 
\end{equation} 

\noindent with $N_c$ -- number of partial waves; $N_p$ -- ensemble dimension; $\textbf{k}_i$ -- wavevectors, computed as independent random variables with PDFs corresponding to the spectrum $S(\textbf{k})$ \eqref{3}; $\omega(\textbf{k}_i)$ -- frequencies \eqref{4}; and $\zeta(\textbf{k}_i)\in[-\pi,\pi]$ -- uniformly distributed random phases. The representation \eqref{5} ensures that, in the limit of $N_c\to\infty$, the Gaussianity of the resulting fields is guaranteed through the Central Limit Theorem, and in the limit of $N_p\to\infty$, the resulting fields converge to the desired statistical properties described above, such as the appropriate autocorrelation function associated with the turbulence spectrum.

From a numerical point of view, we generate $N_c = 10^2$  wavevectors $\textbf{k}$ for each of the $N_p = 1.5 \times 10^5$ realizations of the potential, using the Acceptance-Rejection Method; similarly, the random phases $\zeta(\textbf{k})$ and the associated frequencies $\omega(\textbf{k})$ are constructed. For each of the $N_p$ realizations of the field, the EOMs \eqref{1.1}--\eqref{1.4} are directly solved with a $4^{th}$ order Runge-Kutta method, obtaining $N_p$ trajectories which are then used to compute the asymptotic radial diffusion coefficient \eqref{1.1}. The simulation time is fixed, $t_{\max} = 50$, while the time-step $\delta t$ is chosen a priori in accordance with the parameters of each simulation, assuring that it captures the particle motion:

\begin{align}
	\begin{split}
		\tau_c^{\parallel} &= \frac{\lambda_\parallel}{\sqrt{\langle v_\parallel^2(0) \rangle}} \propto \lambda_\parallel,\\
		\tau_c^{\perp} &= \frac{\lambda_\perp}{\sqrt{\langle v_\perp^2(0) \rangle}} \propto \frac{\lambda_x \rho_i}{A_\Phi \lambda_y a_0},\\
		\delta t &= \frac{1}{5} \min(\tau_c^{\parallel},\tau_c^{\perp}).
	\end{split}
\end{align}

The time-step is fixed for each simulation, and, on average, $\langle \delta t \rangle \approx 0.06$. 

\begin{figure}
	\centering
	\includegraphics[width=0.8\linewidth]{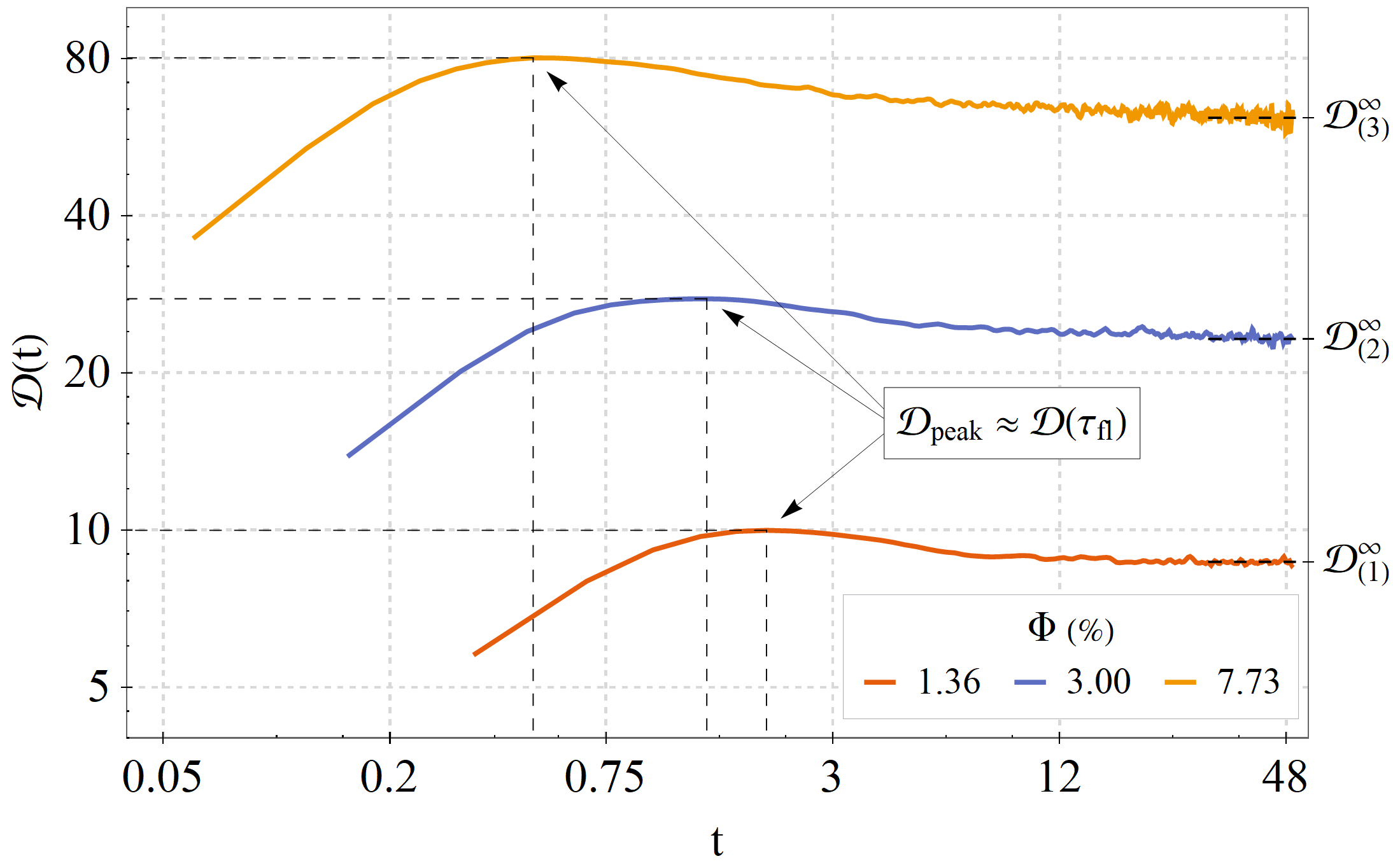}
	\caption{\centering Running radial diffusion profiles for three distinct values of the turbulence amplitude.}
	\label{fig_0}
\end{figure}

Figure \ref{fig_0} shows some typical time-dependent radial diffusion profiles, for three distinct values of the turbulence amplitude $\Phi$; the remaining free parameters are fixed, with $V^*=1,\;\lambda_x=5,\;\lambda_y=5,\;\lambda_z=1.5$. For small times, $t\ll1$, the Lagrangian radial velocities of the particles are $v_x(\textbf{x}(t),t)\approx v_x(\textbf{x}(0),0)$, and the resulting radial diffusion is $\dif(t) \approx \langle v_x^2(0) \rangle t$. The running diffusion reaches its peak around the time-of-flight, $t=\tau_{\mathrm{fl}}$, which is a measure of the time in which the space correlation is lost, and can be approximated as the ratio between the characteristic space-scale of the turbulence and the average amplitude of the velocity field, $\tau_{\mathrm{fl}} = \lambda_x/\sqrt{\langle v_x^2(0)\rangle}$ \cite{palade_pom_ghit_scaling}. This results in a peak diffusion $\dif(\tau_{\mathrm{fl}})\propto \Phi \lambda_x \lambda_y^{-1}$. The diffusive features of the process can be seen in the behavior of the asymptotic diffusion, which saturates at a constant value for $t\gg\tau_{\mathrm{fl}}$.

\subsection{Architecture of Neural Networks}
\label{s1.3}

Neural networks are machine-learning algorithms with a specific architecture that are able to model the relation between input-output variables based on pre-existing data \cite{ANN_basics}. They are adaptable to a vast number of frameworks, systems and tasks, such as pattern recognition/classification \cite{ANN_pattern_recognition}, clustering/categorization \cite{ANN_clustering}, function approximation \cite{ANN_function_approximation, palade_ANN}, prediction \cite{ANN_prediction}, and even dynamic control \cite{ANN_dynamic_control}, working with both discrete and continuous inputs and outputs, and can be applied in a variety of domains. NNs can be classified according to a multitude of characteristics, such as the direction of signal propagation (feed-forward/feed-backward), the number of hidden layers (NNs/deep NNs), the optimization and back-propagation algorithms etc.

\begin{figure}
	\centering
	\includegraphics[width=0.99\linewidth]{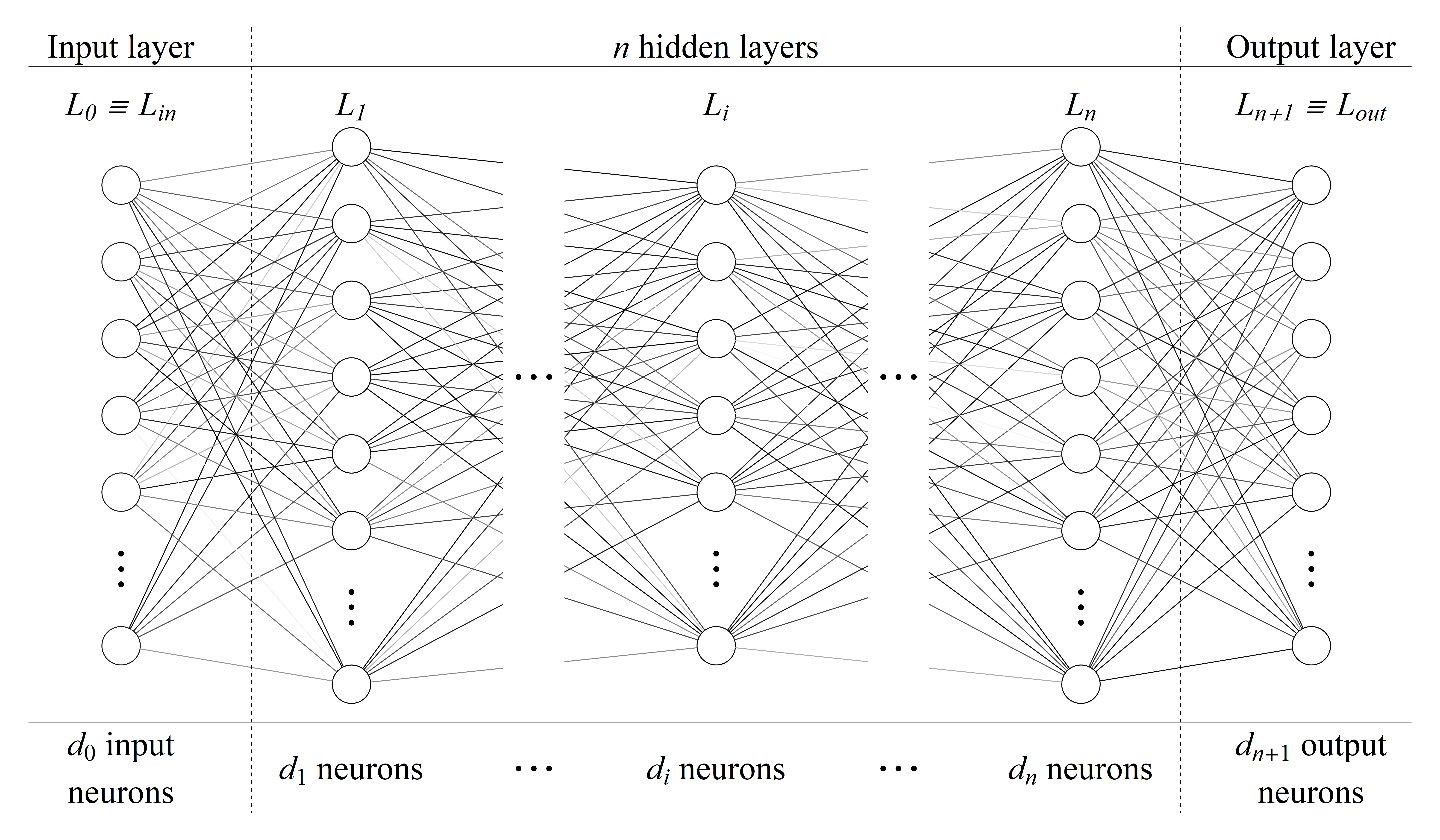} 
	\caption{\justifying General architecture of a NN; the gray level of each connecting line symbolically represents the contribution (weight) of each neuron.}
	\label{fig_1}
\end{figure}

The primary structure of a NN consists of \textit{layers}, which are of three types: one input layer ($L_0\equiv L_{in}$), one output layer ($L_{n+1}\equiv L_{out}$), and between them, multiple hidden layers ($L_i,\;i=\overline{1,n}$). The building blocks of the layers are called \textit{neurons} (based on the slight resemblance between NNs and the biological brain), and each layer can have a different number of neurons (denoted $d_i$). The architecture of a generic NN is schematically represented in Figure \ref{fig_1}. The inputs and outputs of the NN must be data that can be expressed numerically, such that each neuron of the network represents a single numerical value. In this study, the input is the 5-dimensional set of free parameters of the transport model, $\left\{ V^*, \lambda_x, \lambda_y, \lambda_z, \Phi \right\}$, and the output is the single value of the asymptotic radial diffusion, $\dif^{\infty}$.

Equation \eqref{6} describes how $\;\alpha^i_{L_j}$-- the value of the $i^{th}$ neuron of the ${L_j}^{th}$ layer, $\;\alpha^i_{L_j}$, is computed: 

\begin{equation}\label{6}
	\alpha^i_{L_j} = f\left( \sum_{k=1}^{d_{j-1}}\left(w_{L_{j-1}}^{i,k} \cdot \alpha^{i,k}_{L_{j-1}}\right)+\beta^i_{L_j}\right)
\end{equation}

\begin{itemize}
	\item $w_{L_{j-1}}^{i,k}$ corresponds to the \textit{weight} attributed to the value of the $k^{th}$ neuron $\left( \alpha^{i,k}_{L_{j-1}} \right)$ of the previous layer $\left( L_{j-1} \right)$;
	\item the function $f$ denotes the \textit{activation function}, which can be of many types, but is usually chosen either the \texttt{Sigmoid} or the \texttt{tanh} function;
	\item $\beta^i_{L_j}$ corresponds to the \textit{bias} attributed to the $i^{th}$ neuron of layer $L_j$, and it shifts the interval of the activation function's input.
\end{itemize}

Using a matrix notation, the compact form of the equation for all the neurons in a given layer $L_j$ is:

\begin{equation}\label{7}
	\textbf{A}_{L_j} = f\left( \textbf{W}_{L_{j-1}} \textbf{A}_{L_{j-1}}+\textbf{B}_{L_j}\right),
\end{equation}

\noindent where $\textbf{A}_{L_j(-1)}$ is a $d_{j(-1)}\times1$ matrix of all the neuron values of layer $L_{j(-1)}$,\, $\textbf{B}_{L_j}$ is a $d_{j}\times1$ matrix of all the biases corresponding to the neurons of layer $L_{j(-1)}$,\, and $\textbf{W}_{L_{j-1}}$ is a $d_{j}\times d_{j-1}$ matrix of all the weights going from layer $L_{j-1}$ to layer $L_{j}$. Thus, we have $d_j(d_{j-1}+1)$ parameters for the connection between layers $L_{j-1}$ and $L_j$, and the total number of parameters of the NN is:

\begin{equation}\label{8}
	N = \sum_{j=1}^{n+1}d_j(d_{j-1}+1), 
\end{equation}

\noindent where $n$ is the total number of hidden layers.

The NN is trained on preexisting data with the goal of finding the optimal values of the weights and biases in order to minimize the error between the real outputs and its predictions. The training process has a number of components:

\begin{itemize}

\item \textit{Initialization.} Before the training begins, the weights and biases are initialized with random values according to a chosen initialization method. If the parameters start with the same values, or, in the case of the \texttt{tanh} or \texttt{Sigmoid} activation functions -- with absolute values much greater than zero, this regularly slows down or hinders the learning process, leaving the NN in local minima without reaching the global minimum of the function. While the values can be initialized according to various distributions, the most appropriate for our dataset and the activation function used is the normal Xavier method \cite{xavier}

\item \textit{Loss function $\&$ training rounds.} The training process consists of multiple rounds; during a training round, the dataset is split into several batches of a chosen dimension and is fed to the NN. A training round is complete when it went through the whole dataset, and what follows is the backpropagation, and the start of another round. During each round of training, the weights and biases of the NN are adjusted according to an optimization algorithm in order to decrease the error loss function. In this study, we compute the loss function as:

\begin{equation}\label{9}
	\cepsilon_1 = \sqrt{ \frac{\langle\left(\dif^\infty-d^\infty\right)^2\rangle}{\langle\left(\dif^\infty\right)^2\rangle} },
\end{equation}

\noindent where we take the average $\langle \cdot \rangle$ over all the diffusion values of the database, with $\dif^\infty$ -- real value of the training data (TD) or the validation data (VD), and $d^\infty$ -- value predicted by the network.

\item \textit{Optimization.} The most common optimization method used is the Stochastic Gradient Descent (SGD) \cite{SGD}; in each iteration, a batch is randomly selected, and the gradient of the loss function with respect to the weights is computed; the weights are then updated in the opposite direction of the gradient, to reduce the loss. In this work, we employ a widely used variant of SGD -- the Adaptive Moment Estimation (ADAM) method \cite{ADAM}, which is an adaptive-learning-rate method.

\end{itemize}

\section{Numerical details of the NN and the database}
\label{s2}

The free parameters of the model described in Section \ref{s1} can be divided into three categories: tokamak parameters, plasma equilibrium profile parameters, and turbulence parameters. Starting with the latter, these variables characterize the turbulence profile of the electric potential, such as the correlation lengths $\lambda_x$, $\lambda_y$ and $\lambda_z$, the position of the maxima of the $k_y$ spectrum, $\pm k_0$, and the turbulence strength $\Phi$. 

The scaled diamagnetic drift velocity, $V^*$, the ion temperature, the particle density, and the pressure gradients are examples of plasma equilibrium profile parameters. The tokamak parameters characterize the specific device that's to be studied; these include the major and minor radii, $R_0$ and $a_0$, the intensity of the magnetic field, $B_0$, and, implicitly, the magnitude of the grad-B and curvature drifts. Due to the large number of variables, we choose to restrict our research to only one tokamak device, the ASDEX Upgrade (AUG); this sets fixed values for all the tokamak parameters.

We can reproduce the same characteristics of the $k_y$ spectrum (eq. \eqref{3}), as well as the corresponding asymptotic diffusion, by fixing $k_0$ and only varying the correlation length $\lambda_y$. Hence, the remaining free parameters of the model (which will be input parameters for our NN) are: the diamagnetic velocity $V^*$, the three correlation lengths $\lambda_x, \;\lambda_y, $ and $\lambda_z$, and the turbulence strength $\Phi$. Table \ref{table_1} further details the fixed values and value intervals chosen for the model's parameters. 

\begin{center}
	\begin{tabular}{ccclc|cccll} \toprule[0.03cm]\toprule[0.03cm]
		
		{$a_0$} & {\textit{minor radius}} 				& {0.65}	& {$m$}	& { }  	& { }  &	{$\lambda_x$} & {\textit{$x$-correlation length}}  & {$\left[1.0,10.0 \right]$}  & {$\rho_i$}\\
		{$R_0$} & {\textit{major radius}} 				& {1.5} 	& {$m$}	& { }  	& { }  &	{$\lambda_y$} & {\textit{$y$-correlation length}}  & {$\left[1.5,10.0 \right]$} & {$\rho_i$}\\
		{$B_0$} & {\textit{magnetic field strength}}  	& {2.5}  & {$T$}	& { }  	& { }  &	{$\lambda_z$} & {\textit{$z$-correlation length}}  & {$\left[0.5,2.5\right]$} & {$\rho_i$}\\ 
		{$T_i$} & {\textit{ion temperature}}  			& {1.0}  & {$keV$}	& { }  	& { }  &	{$V^*$} & {\textit{$\;\;\;$diamagnetic drift velocity$\;\;\;$}}  & {$\left[0.0,2.0 \right] $} & {$v_{th} \rho_i/a_0$}\\ 
		{$k_0$} & {\textit{$\;\;\;$auxiliary $S(\textbf{k})$ parameter$\;\;\;$}}  & {0.1}& {$\rho_i^{-1}$}	& { }  	& { }  &	{$\Phi$} & {\textit{turbulence strength}}  & {$\left[0.0, 10.0 \right]$} &{$\%$} \\\midrule
		{$A$}  	& {\textit{mass number}} 	& {1}  		& {}	& { }   & { } &	{$N_c$} & {\textit{no. of partial waves}}  & {$\;\;\;\;\,10^2$} & {}\\ 
		{$Z$}  	& {\textit{charge number }} & {1} 		& {} 	& { }	& { }  &	{$N_p$} & {\textit{ensemble dimension}}  & {$1.5 \times 10^5$} & {}\\ \bottomrule
	
	\end{tabular}
	\captionof{table}{\centering Values of the fixed and free parameters of the model, in accordance with \cite{AUG_parameters_1, AUG_parameters_2}.}
	\label{table_1}
\end{center}

\begin{figure}
	\centering
	\includegraphics[width=0.8\linewidth]{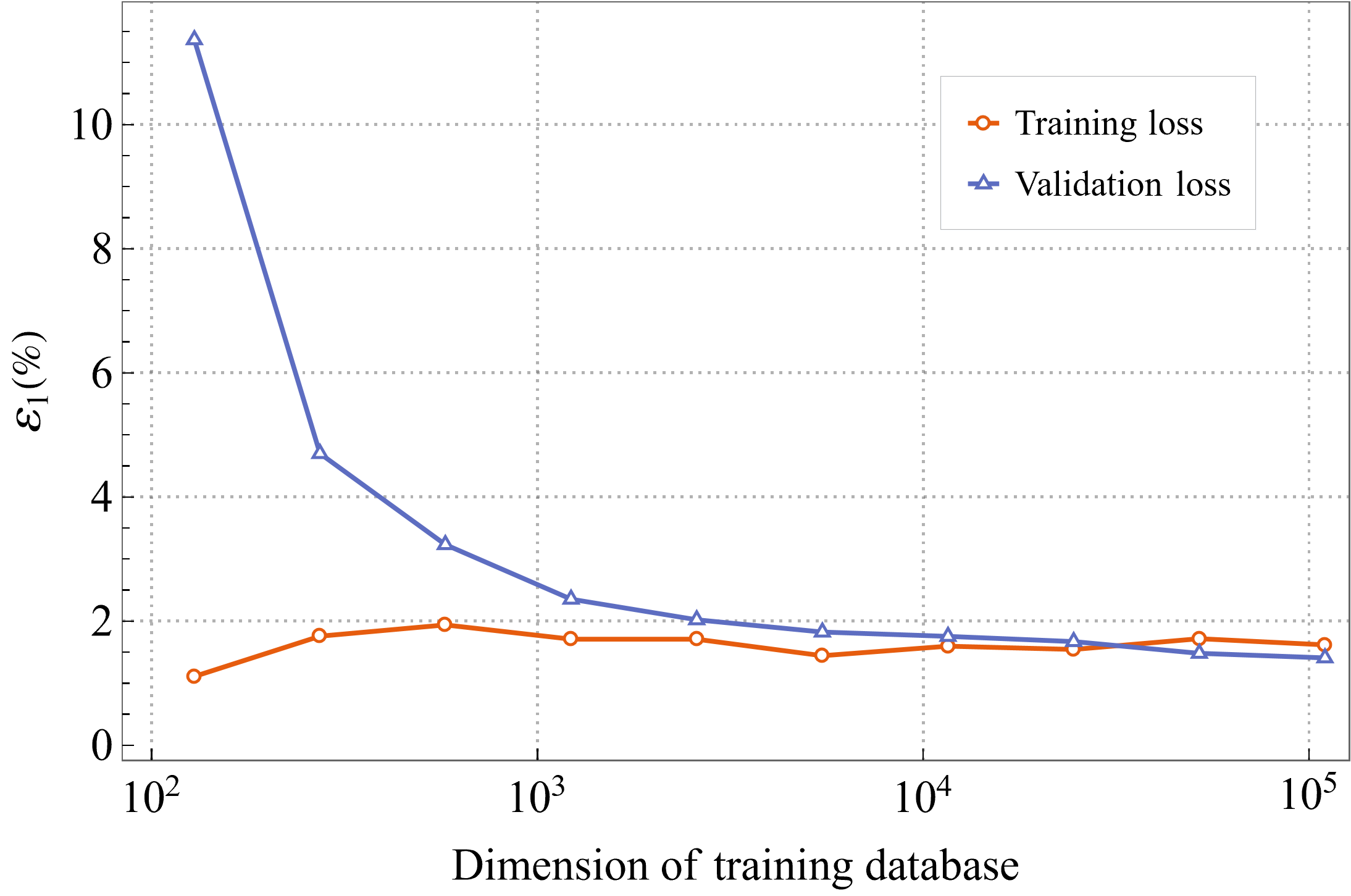}
	\caption{\justifying Dependence between the training (orange circles) and validation (blue triangles) error loss function $\cepsilon_1$ and the dimension of the TD.}
	\label{fig_2}
\end{figure}

Due to the random initialization of the weights and biases, no two trained nets result in the same exact configuration. While a point as close as possible to the global minimum should be reached for all the training processes, the paths taken to achieve it differ. Moreover, we can never be certain that the global minimum \textit{is} reached. Therefore, each trained net yields slightly different predictions, and while the error of either one is minimal, it isn't negligible. A viable workaround for this issue is training multiple nets and constructing the predictions as averages; in literature, this is coined as an \textit{ensemble neural network model} \cite{ensemble_ann}. The primary reason this linear function of nets yields better results than any of the constituent NNs is in virtue of the Central Limit Theorem, as the errors and biases of the individual networks tend to cancel out and, for large enough ensembles, approach the true value of the asymptotic diffusion. In this work, we trained multiple NNs with architectures corresponding to combinations of 2, 3, and 4 layers, and 15, 30, 45 and 60 neurons (not including the 5 inputs and 1 output); this leads to an ensemble of 12 NNs with varied architectures, and the results presented below are averages of their predictions.

In order to improve the accuracy of the network and to speed up the training process, we normalize the input and output data prior to feeding it to the NN. This pre-processing step is crucial if the activation function used is the \texttt{Sigmoid}, \texttt{tanh}, or similar functions, as their outputs are bounded between 0 and $-1$, or $\pm1$. The most common way to standardize the data is rescaling it with a mean of $0$ and a variance of $1$. However, for the dataset at hand, this leaves a lot of output values outside the range of the \texttt{tanh} activation function. Therefore, we rescale the inputs/outputs to be bounded between $\pm 1$, by applying a linear transformation:

\begin{figure}
	\centering
	\includegraphics[width=0.8\linewidth]{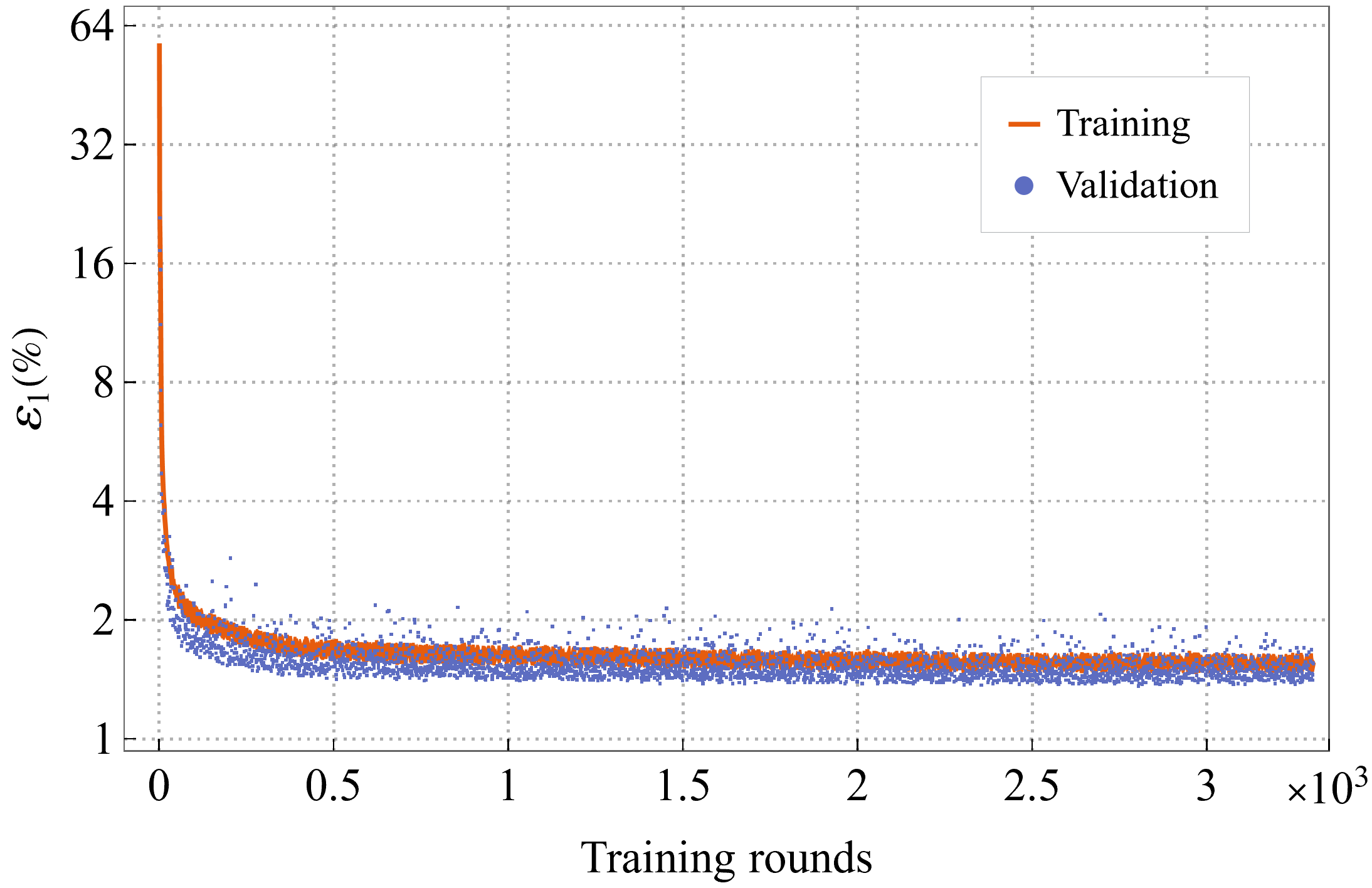}
	\caption{\justifying Typical evolution of the percentage error loss function $\cepsilon_1$ during the training of one of the 12 NNs, consisting of 2 hidden layers with 20 and 10 neurons, respectively.}
	\label{fig_6}
\end{figure}

\begin{equation}\label{10}
	\begin{split}
		\bar{X}&=2\frac{X-m}{M}-1, \\ 
		m&=\min\left[X\right],\\
		M&=\max\left[X\right]-m,
	\end{split}
\end{equation}

\noindent with $X$ representing either the training inputs, or the training outputs of the NN. This normalization assures that the activation function is not linear for the given parameter intervals, and that the output values of the NN are, indeed, bounded between $\pm 1$.

Regarding the dimension of the TD, we conclude that a training set consisting of $N = 10^5$ datapoints should suffice for the purposes of the present work, as the accuracy of the NN tends to converge at $N \propto 10^4$ datapoints. The dependence of the final percentage loss error $\cepsilon_1$ on the dimension of the dataset can be seen in Figure \ref{fig_2}. We chose to distribute the parameters uniformly inside a 5-dimensional hypercube, assigning 10 values to each of the five variables. The decision to use a uniform grid for the parameters (rather than sampling random values within the intervals of interest) was motivated by the intention to compare the performance of the NN with that of a simple interpolation, which works best when the input parameters are distributed on an equidistant grid.

In addition to the $10^5$ values obtained by test-particle simulations, we complete the TD with $10^4$ analytic values, for $\Phi=0$ and for which the asymptotic diffusion is $\dif^\infty=0$. For the validation and testing of the NN, we construct a validation dataset consisting of $2 \times 10^4$ randomly generated points inside the $5-$dimensional hypercube of parameters. We complete this testing database with: points for which we vary each parameter individually (keeping the other 4 parameters constant); points for which we vary the parameters in pairs (keeping 3 parameters constant), and so on. The structural details of the whole database are summarized in Table \ref{table_2}.

\begin{center}
\begin{tabular}{rl|c|c} \toprule[0.03cm]\toprule[0.03cm]
{Trai}& {\hspace{-0.7ex}ning$\;\;$} 		& {$\;\;$1 subset $ \times\; 10^5$ + $10^4$ analytic$\;\;$}	& {$\;\;$$1.10\times 10^5\;$ values$\;\;$}\\\midrule
{Valid}& {\hspace{-0.7ex}ation$\;\;$} 	& {$\;\;$$2\times10^4$ randomly generated$\;\;$} 			& {$\;\;$$2.00\times 10^4\;$ values$\;\;$}\\\midrule
\multirow{4}{*}{\begin{sideways}T e s t i n g\end{sideways}\;\;\;} 
	& {$V_1$} 					& {\hspace{1ex}$\;\;$ 5 subsets $\times \;100^1$}					& {$\;\;$$5.00\times10^2\;$ values $\;\;$}	\\
{} 	& {$V_2$} 					& {\hspace{-2.5pt}$\;\;$10 subsets $\times\;25^2$} 							& {$\;\;$$6.25\times10^3\;$ values$\;\;$} 	\\
{}	& {$V_3$}  					& {$\;\;$10 subsets $\times\; 11^3$ } 							& {$\;\;$$1.00\times10^4\;$ values$\;\;$}  	\\ 
{}	& {$V_4$}  					& {$\;\;$5 subsets $\times \;6^4$ } 							& {$\;\;$$6.48\times 10^3\;$ values$\;\;$} 	\\ \bottomrule
\multicolumn{4}{r}															{\textbf{Total:}$\;\;$$\;1.53 \times 10^5\;$   values$\;\;\,$} \\ \bottomrule

\end{tabular}\\
	\captionof{table}{\centering Dimensions and structure of the training, validation and testing databases; the testing sets $V_{1-4}$ each have 1, 2, 3 or 4 free parameters varied, with the rest of the variables fixed, and the subsets represent combinations of the five input parameters.}
	\label{table_2}
\end{center}

\section{Results}
\label{s3}

\subsection{Training process and prediction accuracy of the NN}
\label{s3.1}

\begin{figure}\label{fig_7}
	\subfloat[\hspace{-15pt}\label{fig_7a}]{
		\includegraphics[width=.475\linewidth]{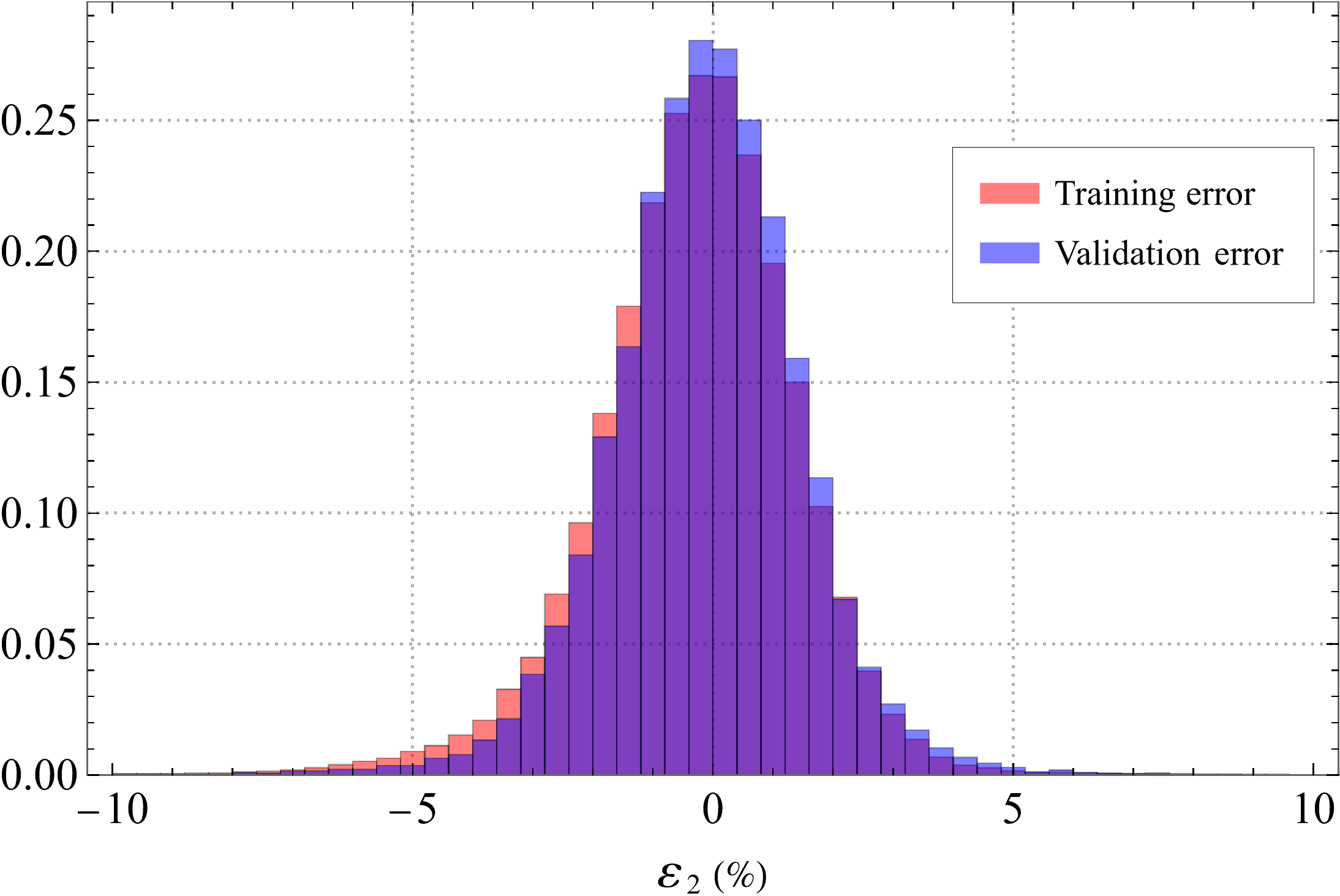}%
	}	
	\subfloat[\hspace{-31pt}\label{fig_7b}]{
		\includegraphics[width=.515\linewidth]{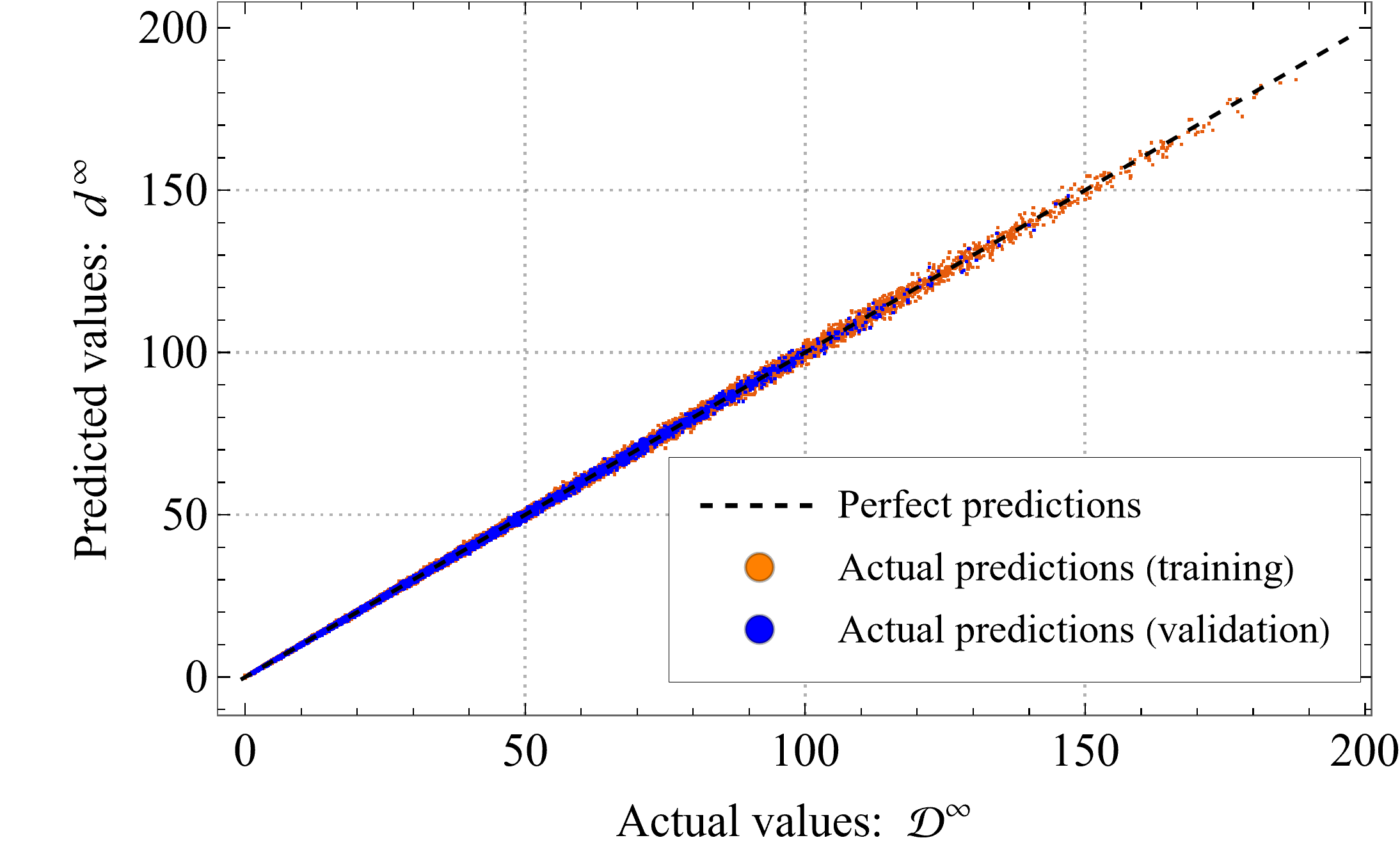}%
	}
	\caption{\justifying \textit{(Left)} Relative error $\cepsilon_2$ between the actual $\left(\dif^\infty\right)$ and the predicted values of the asymptotic diffusion $\left(d^\infty \right)$, for the training (orange, back) and for the VD (blue, front). \textit{(Right)} Dependence between the actual and the predicted values of the asymptotic diffusion, for the TD (orange) and for the VD (blue).}
\end{figure}

Let us take a closer look at the training process and the convergence properties of the aforementioned NN ensemble. The training stops when a criterion is met, such as the relative error between the loss of two consecutive rounds being less than a set value, or obtaining the same error for an arbitrary number of rounds, both of which indicate saturation. In this work, we used both criteria, demanding that the error function is constant \textit{and} below the threshold for $10^3$ training rounds. On average, the NNs need around $2.5\times10^3$ training rounds before convergence is achieved, which is equivalent to $\sim \hspace{-3pt} 2$ hours of elapsed time on a personal CPU. A typical evolution of the error loss function $\cepsilon_1$ during the training is presented in Figure \ref{fig_6}; the net trained in this example is structured in 2 hidden layers, with 20 and 10 neurons, respectively. We see that the the NN saturates at a constant value around the $2000^{th}$ training round; the constant decrease and the saturation indicate that the net has reached a point close to the global minimum, and is not overfitting the data.

\begin{figure*}\label{fig_8}
	\renewcommand\tabularxcolumn[1]{m{#1}}
	\setkeys{Gin}{width=\linewidth, 
		height=0.9\linewidth,
		keepaspectratio}
	\begin{tabularx}{\linewidth}{@{} XX @{}}
		
		\subfloat[\hspace{-23pt}\label{fig_8a}]{\includegraphics{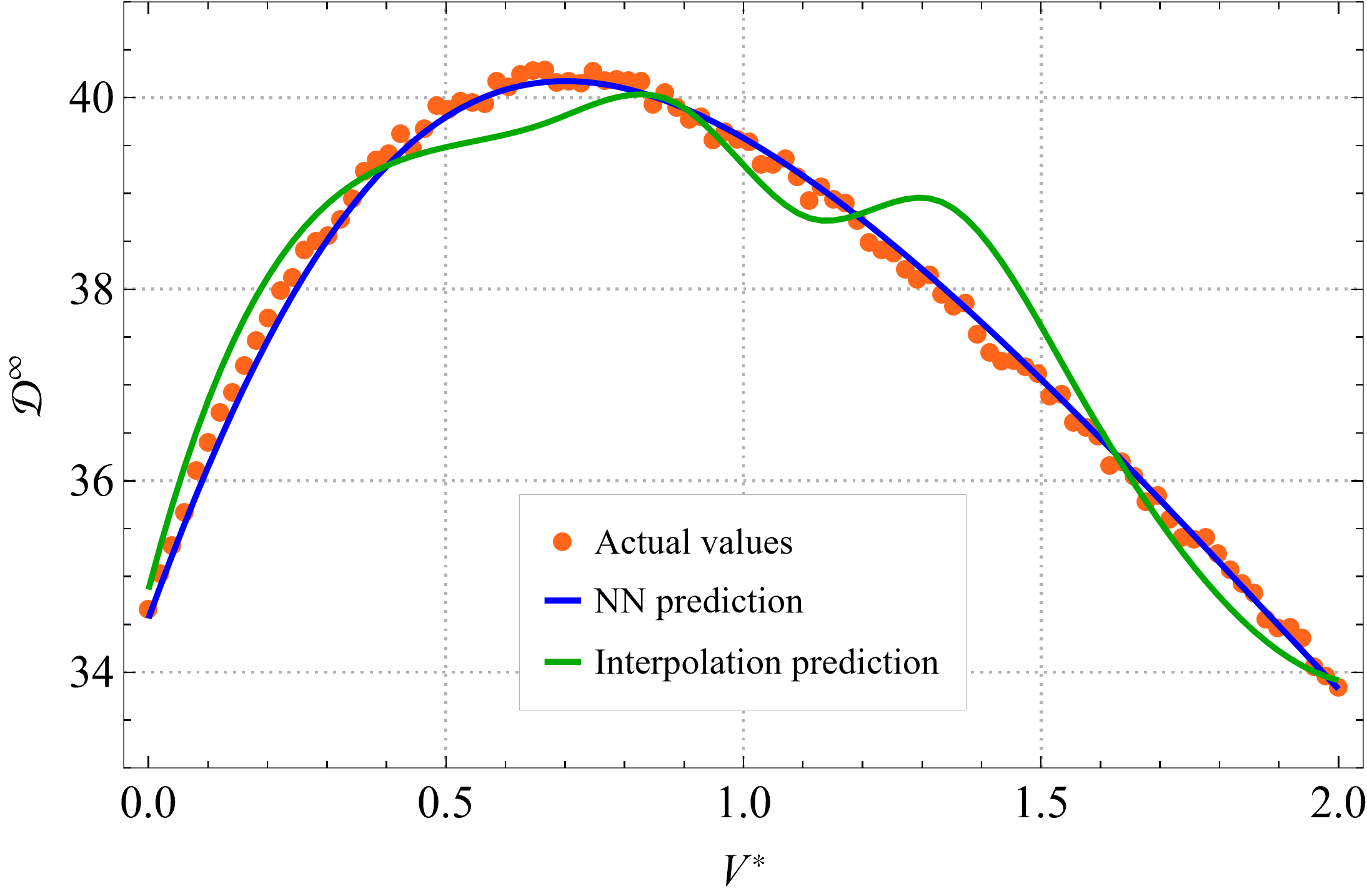}}
		
		\subfloat[\hspace{-23pt}\label{fig_8b}]{\includegraphics{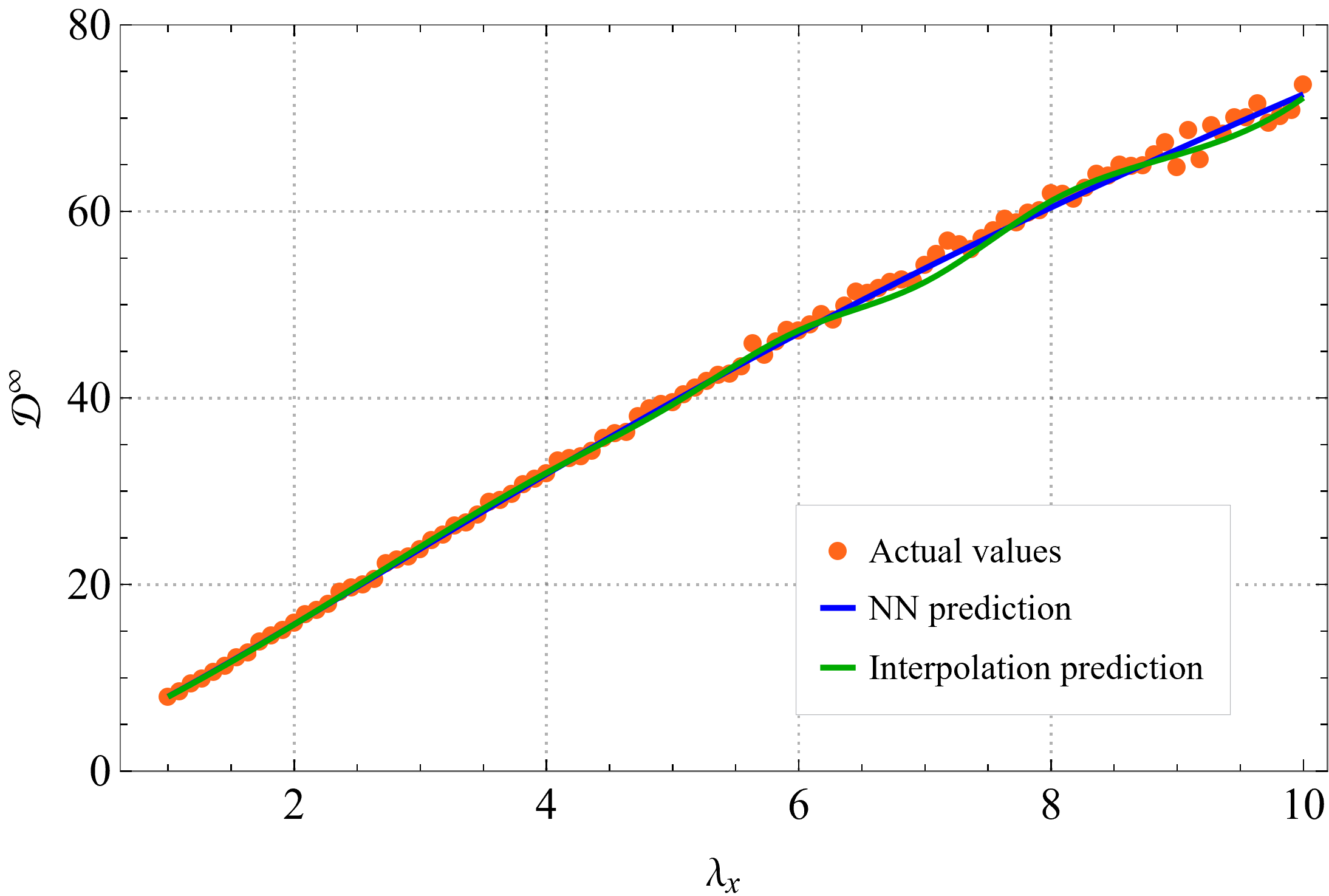}}
		
		\subfloat[\hspace{-23pt}\label{fig_8c}]{\includegraphics{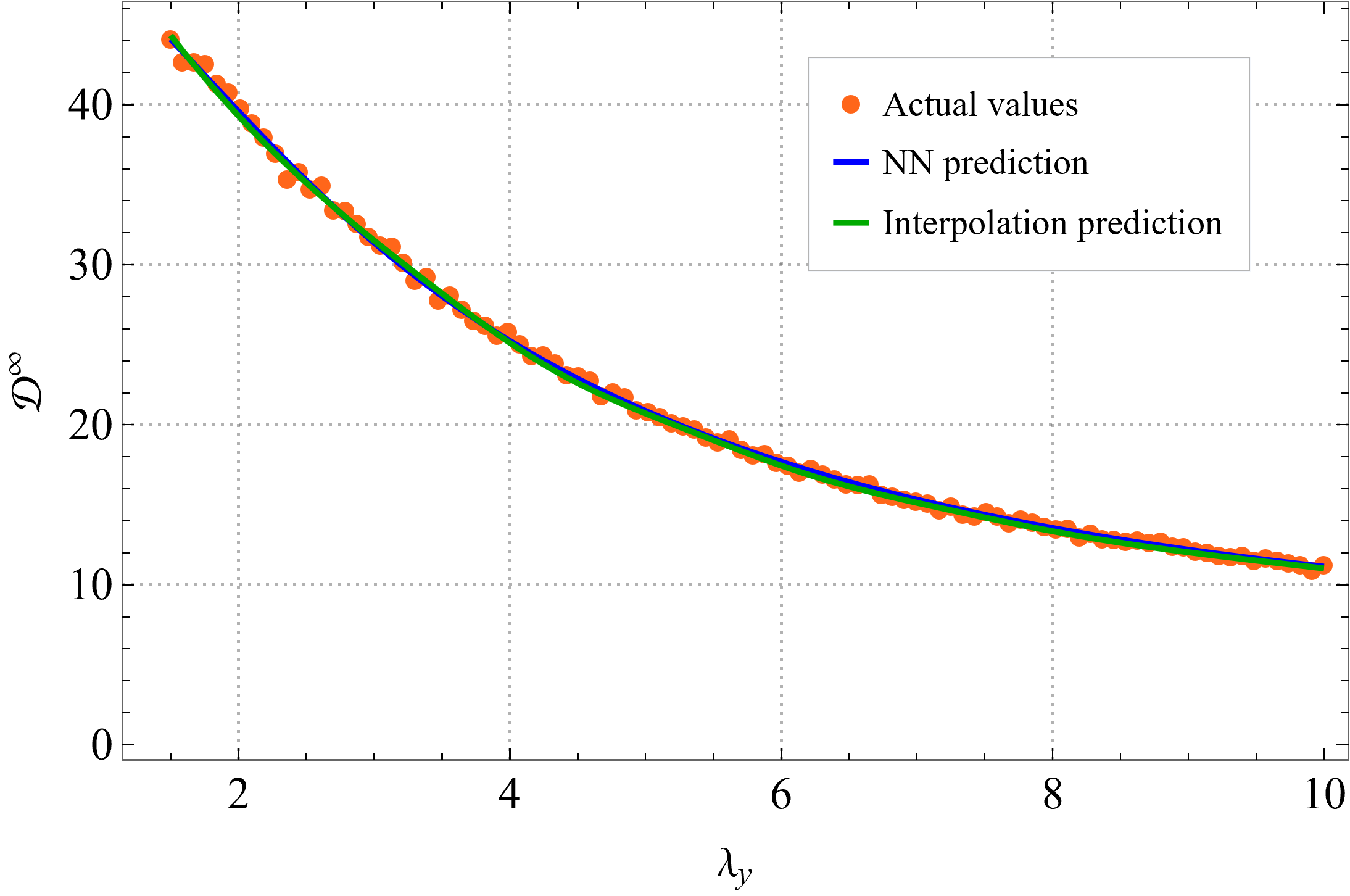}}
		&
		
		\subfloat[\hspace{-23pt}\label{fig_8d}]{\includegraphics{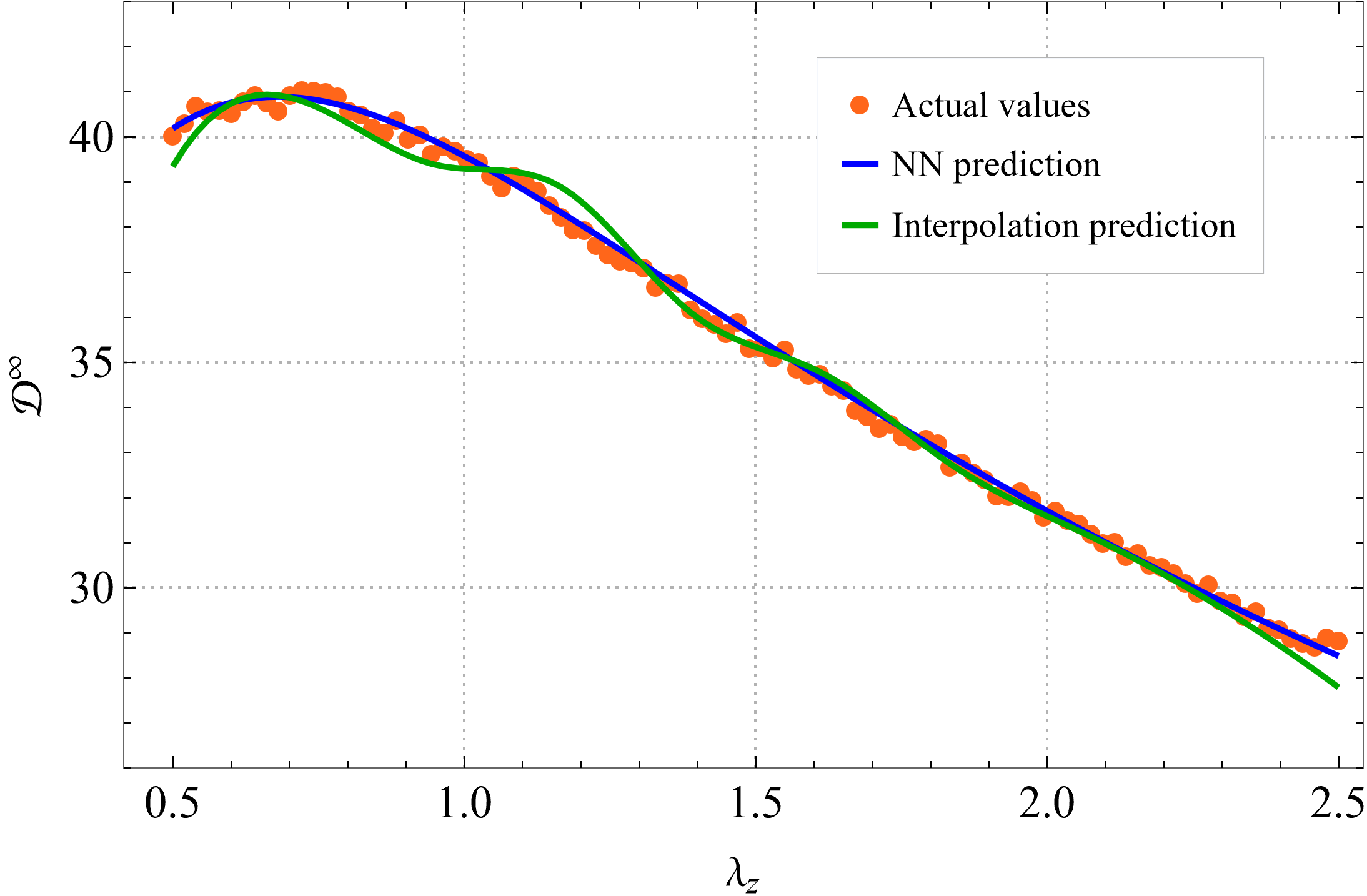}}
		
		\subfloat[\hspace{-23pt}\label{fig_8e}]{\includegraphics{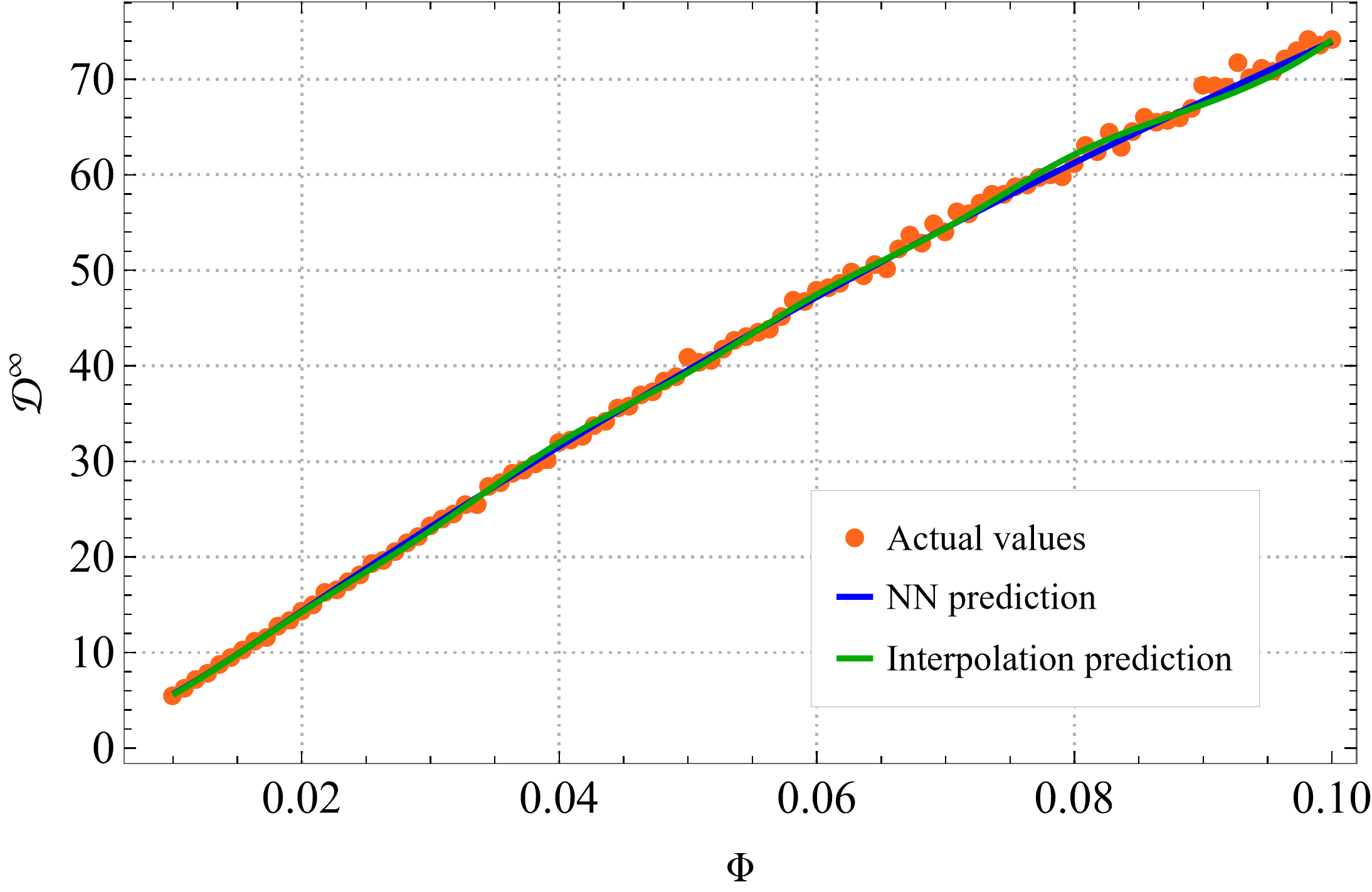}}
		\color{white}
		\vspace{11.5ex}
		\color{black}
		
	\end{tabularx}
	\caption{ \justifying Dependence of the asymptotic radial diffusion on the free parameters of the model, each one varied individually; all figures feature the actual data (orange points), the NN predictions (blue lines), and the interpolation predictions (green lines).}
\end{figure*}

Figure \ref{fig_7a} shows the relative error distribution between the actual values of the asymptotic diffusion $\left(\dif^\infty\right)$ and the values predicted by the NN $\left(d^\infty \right)$:

 $$\cepsilon_2 = 1- d^\infty/\dif^\infty.$$

We see that the error is centered around $0\%$, with a standard deviation of $\sim \hspace{-3pt} 2\%$ for the TD, and $\sim \hspace{-3pt}1.5\%$ for the VD. In Figure \ref{fig_7b}, we plot the dependence between the real and the predicted values of the diffusion. We see that the error is evenly distributed on the studied intervals and is not clustered in any specific regions; this, together with the symmetry of the error distribution of Figure \ref{fig_7a}, indicates that the output of the NN is not biased (i.e. systematically sub- or over-estimating the diffusion), nor does it fail for specific values or intervals of the parameters. 

\begin{figure}
	
	\begin{minipage}{.5\linewidth}
		\centering
		\subfloat[\hspace{55pt}\label{fig_9a}]{\includegraphics[width=0.99\linewidth]{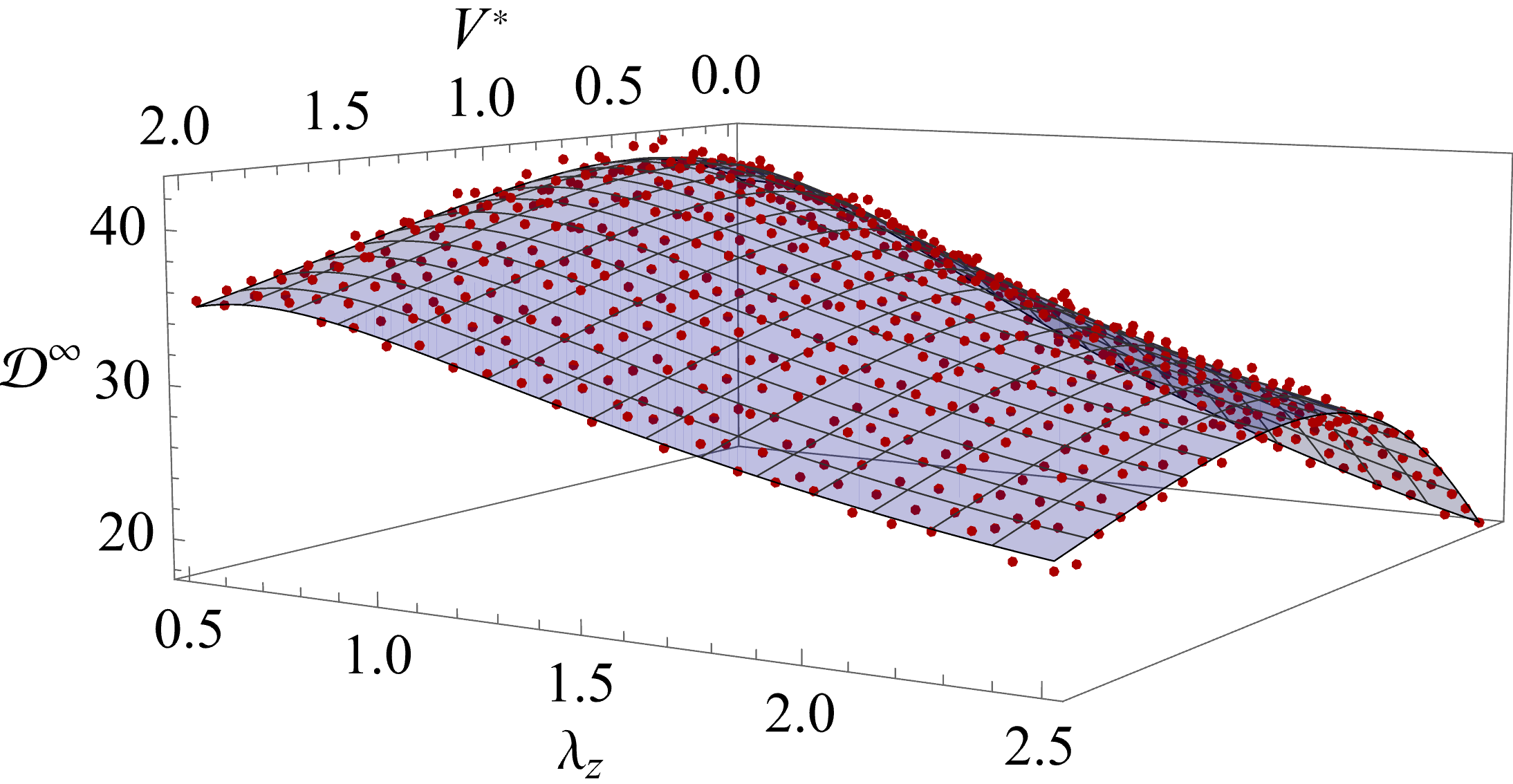}}
	\end{minipage}%
	\begin{minipage}{.5\linewidth}
		\centering
		\subfloat[\hspace{-59pt}\label{fig_9b}]{\includegraphics[width=0.99\linewidth]{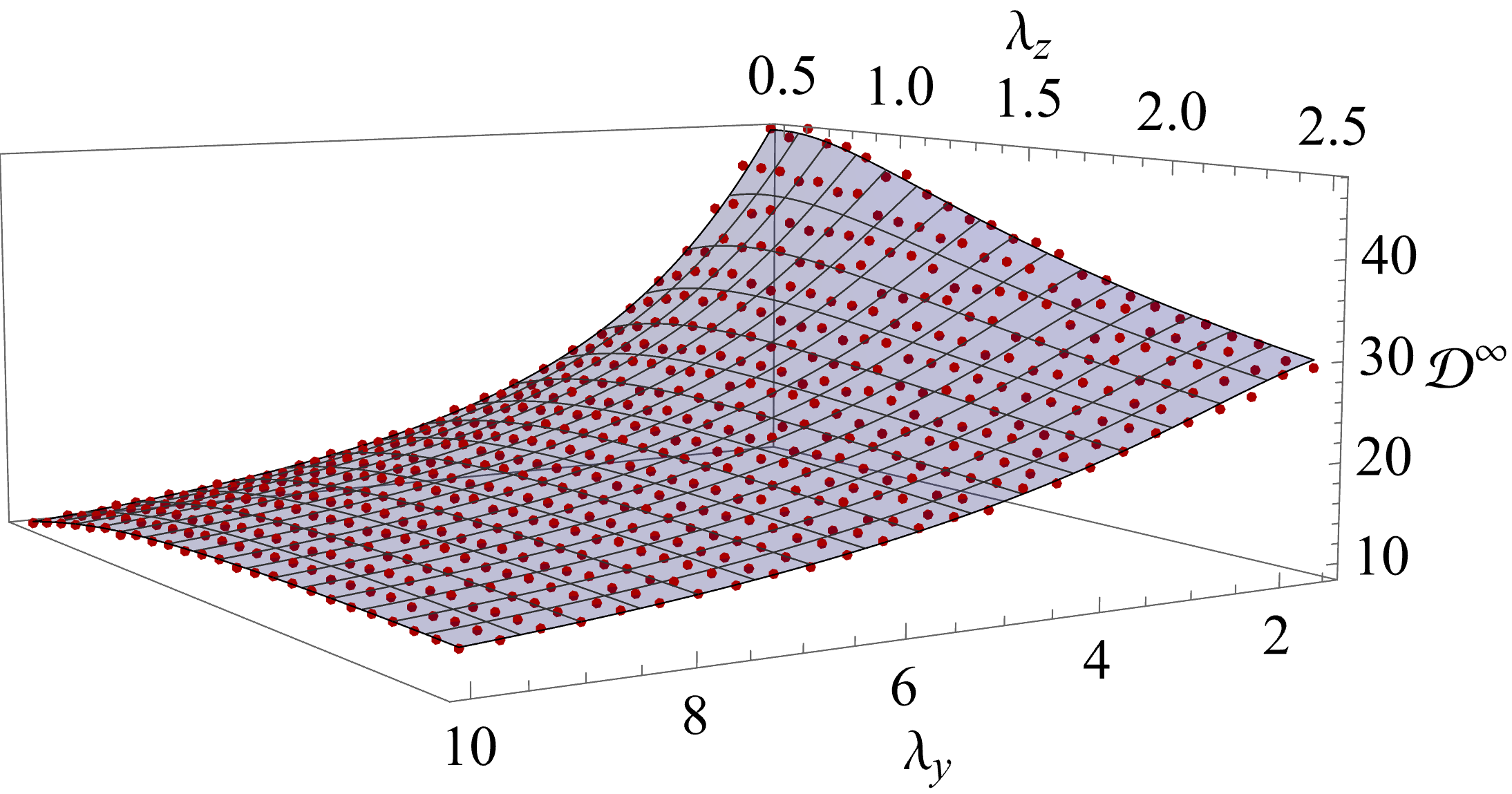}}
	\end{minipage}\par\medskip
	\centering
	\vspace{-12ex}
	\hspace{-3.ex}
	\subfloat[\hspace{-28pt}\label{fig_9c}]{\includegraphics[width=.495\linewidth]{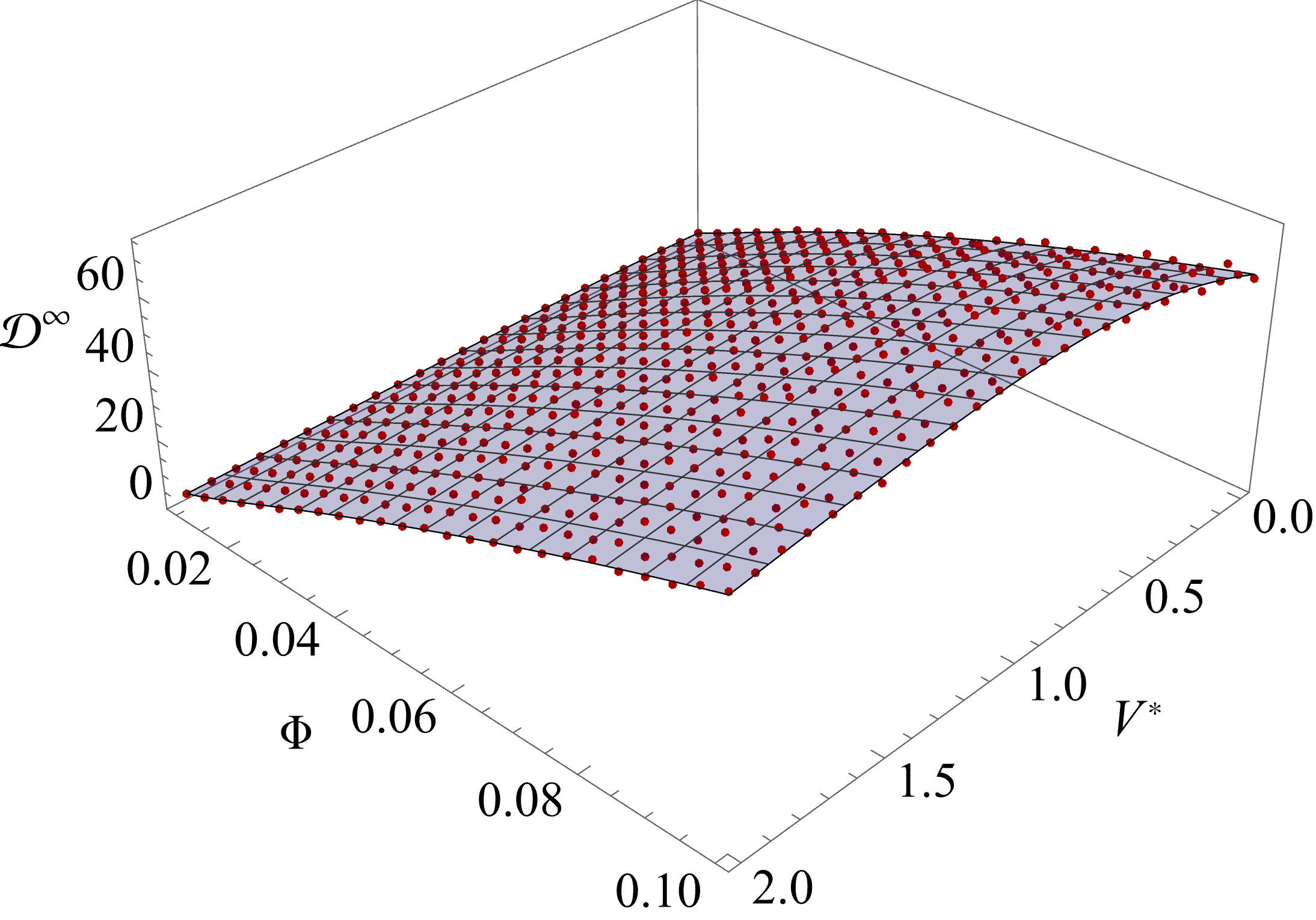}}

	\caption{	\centering Real data (red points) and NN predictions (blue surfaces) of the dependence between the radial asymptotic diffusion $\dif^{\infty}$ and pairs of two parameters $\left(\left\{ \lambda_z, V^* \right\} - \mathrm{(a)}, \; \left\{ \lambda_y, \lambda_z \right\} - \mathrm{(b)},\; \left\{ \Phi, V^* \right\} - \mathrm{(c)} \right)$.}
	\label{fig_9}
\end{figure}

Further on, we inquire the predicting capabilities of the trained network ensemble by looking at the testing datasets $V_{1-4}$ (see Table \ref{table_2}). When one or more parameters are not varied, they are fixed at the following base values, around the middle of the intervals used: $V^*=1,\;\lambda_x=5,\;\lambda_y=5,\;\lambda_z=1.5,\;\Phi=0.05$. In \Cref{fig_8a,fig_8b,fig_8c,fig_8d,fig_8e}, we compare the real values of the asymptotic radial diffusion (orange points) with the ones predicted by the NN (blue lines), varying each of the 5 parameters individually; in these figures, the green lines correspond to the predictions of an interpolation, which are discussed further in Section \ref{s3.2}. In \Cref{fig_9a,fig_9b,fig_9c}, we plot the real values of the diffusion (red points) over the surfaces of the NN predictions, with two parameters varied at a time, and the rest fixed. We note that although the dependencies of the asymptotic diffusion are fairly simple, the predictions made by the NN ensemble are unexpectedly close to the real values; more details on the accuracy of the predictions are presented in Table \ref{table_3}.

\begin{center}
	\begin{tabularx}{\textwidth}{c *5{>{\Centering}X}}
		\toprule \toprule
		{}& \multicolumn{1}{c}{$\;\mu=\langle \cepsilon_2\rangle\;$} & \multicolumn{1}{c}{$\; \sigma=\sqrt{ \langle \left(\cepsilon_2-\mu\right)^2 \rangle } \;$}& \multicolumn{1}{c}{$\; \sqrt{\langle \left( d^{\infty} \right)^2\rangle/\langle \left( \dif^{\infty} \right)^2\rangle} \;$}& \multicolumn{1}{c}{$\;\%$ of $\lvert\cepsilon_2 \rvert < 10\%\;$} \\ 
		\midrule
		{TD}		& {-0.28\%} & {1.74\%} 	& {0.999} & {99.90\%} \\
		{VD}		& {-0.04\%} & {1.45\%}	& {0.999} & {99.98\%} \\
		{$V_{1-4}$}	& {-0.15\%} & {1.59\%}  & {0.997} & {99.96\%} \\
		\bottomrule
	\end{tabularx}
	\captionof{table}{\justifying Statistics of the percentage error $\cepsilon_2$ distribution, for the predictions of the NN trained on the TD.}
	\label{table_3}
\end{center}

 some fluctuations of a numerical nature in the results.

\subsection{Predicting turbulent transport using NNs vs interpolation}
\label{s3.2}

There is a question of whether the predictive tasks assigned to the NN could be more easily achieved through a simple interpolation of the TD. The latter has been constructed using the built-in \texttt{Interpolation} function of \textit{Wolfram Mathematica $\mathrm{v11.3}$} \cite{mathematica}, using \texttt{Method} $\to$ \texttt{"Spline"}. After constructing the \texttt{InterpolatingFunction}, we look at its predictions on the testing dataset $V_1$ (the green squares of \Cref{fig_8a,fig_8b,fig_8c,fig_8d,fig_8e}). While there aren't notable differences in the predictions of the NN ensemble and those of the \texttt{InterpolatingFunction} for the monotonic dependencies of the asymptotic diffusion with $\lambda_x$, $\Phi$, or $\lambda_y$, the \texttt{InterpolatingFunction} shows some fluctuations for the more complex dependencies with $V^*$ and $\lambda_z$.  Overall, the \texttt{InterpolatingFunction} predictions are intrinsically exact on the TD, and sufficiently close to the real values on the VD and $V_{1-4}$. We also note that constructing the \texttt{InterpolatingFunction} takes mere seconds, whereas training the NN ensemble requires close to 24 hours. 

The next step is looking at the extrapolating properties of the two. For this, we truncate the TD, removing the two outermost values for each of the 5 parameters; this results in $\sim \hspace{-3pt} 3.2 \times 10^4$ values remaining for the training. A comparison between the performance of the NN and the performance of the \texttt{InterpolatingFunction} (Int) is detailed in Table \ref{table_4}.

One noteworthy advantage of NNs is that their training doesn't require the input parameters to be distributed on an equidistant grid, which makes it easier to add further values to the initial TD. Therefore, we conclude that the NN ensemble is slightly more accurate and convenient within the training data range, and significantly so outside it.

\begin{center}
	\begin{tabularx}{\textwidth}{c *9{>{\Centering}X}}
		\toprule \toprule
		{}& \multicolumn{2}{c}{$\;\mu=\langle \cepsilon_2\rangle\;$} & \multicolumn{2}{c}{$\; \sigma=\sqrt{ \langle \left(\cepsilon_2-\mu\right)^2 \rangle } \;$}& \multicolumn{2}{c}{$\; \sqrt{\langle \left( \dif^{\infty} \right)^2\rangle/\langle \left( d^{\infty} \right)^2\rangle} \;$}& \multicolumn{2}{c}{$\;\%$ of $\lvert\cepsilon_2 \rvert < 10\%\;$} \\ 
		\cmidrule(lr){2-3} \cmidrule(l){4-5}\cmidrule(l){6-7} \cmidrule(l){8-9}
		{}  & {NN}      & {Int}   & {NN}     & {Int}   & {NN}   & {Int} & {NN} & {Int}\\ \midrule
		{TD}& {-0.82\%} & {3.99\%}& {5.87\%} & {5883.16\%}& {1.002} & {40.178}& {94.02\%} & {59.37\%}\\
		{VD}& {-0.46\%} & {-2.98\%}& {2.35\%} & {213.11\%}& {1.005} & {1.879}& {99.41\%} & {80.13\%}\\
		{$V_{1-4}$}& {-0.95\%} & {-4.82\%}& {4.32\%} & {1889.29\%}& {1.004} & {22.311}& {96.33\%} & {18.37\%}\\
		\bottomrule
	\end{tabularx}
	\captionof{table}{\centering Statistics of the percentage error $\cepsilon_2$ distribution, for the predictions of the NN in comparison to the \texttt{InterpolatingFunction}, both trained on the truncated dataset.}
	\label{table_4}
\end{center}

\subsection{A word on the physical mechanisms}
\label{s3.3}

The main goal of this work is to give a detailed description of the methodology for the development of NNs for predicting turbulent transport. Evidently, this goes hand in hand with the development of an associated database; the latter was constructed with the aid of test-particle numerical simulations, in the frame-work of a statistical approach on a simplified transport model. 

Although not directly related to the NN, it is useful to have an understanding of the physical processes involved in the transport, and of how various plasma and turbulence properties impact it. This might be significant because, while not employed in this study, having previous knowledge of the gross dependence between inputs and outputs could enable us to develop more reliable databases (perhaps non-uniform grids), or even use specific activation functions in the NNs. The mathematical and physical factors beyond the dependence of the output, $\dif^\infty$, on each of the input parameters, $\left\{ V^*, \lambda_x, \lambda_y, \lambda_z, \Phi \right\}$, will be individually examined. 

First, let us take a look at the transport model of eqs. \eqref{1.1} -- \eqref{1.4}. Perhaps the simplest limit of that model is when the particles considered are very cold ($T_i=0$, which implies $v_{\parallel}=0, \mu=0$), the turbulent field is frozen ($\omega = 0$, or flat pressure profiles), and the scale of the parallel fluctuations of the turbulence is very large ($\lambda_z \to \infty$). In this limit, it is straightforward to see that the only term remaining is the $\textbf{E}\times \textbf{B}$ drift in a time-independent potential, which is equivalent with a 2D Hamiltonian system. Consequently, all the trajectories are closed (i.e. trapped particles), and the running diffusion coefficient has an algebraic decay to 0 with time \cite{palade_pom_ghit_scaling}; this is due to the fact that the particles remain correlated at any latter time.

The next level of complexity is considering non-zero temperatures and a finite parallel correlation length. This implies that the particles experience parallel velocities (even in the absence of a parallel acceleration) which, in the wave decomposition of the turbulent field, makes 

$$\sin(\textbf{k} \cdot \textbf{x}(t)+\zeta) \to \sin(k_x x(t)+k_y y(t) +k_z z(t)+\zeta) \to  \sin(k_x x(t)+k_y y(t) +k_z v_{\parallel} t+\zeta).$$

Thus, a frequency-like term ($k_z v_{\parallel}$) is acquired by the field, which makes it time-dependent. The Hamiltonian characteristic of the dynamics is broken and the trajectories  are not closed anymore; they are able to explore various regions in the perpendicular plane, be scattered by the potential landscape, and exhibit finite (non-zero) diffusion. In essence, this is a decorrelation mechanism induced through the motion in the landscape of parallel fluctuations of the turbulence; an associated decorrelation time would be $\tau_c = \lambda_z/v_{\parallel}$. It is expected that the parallel acceleration will contribute to this decorrelation by introducing more stochasticity in the parallel velocities, yet it is difficult to infer how much more.

Another decorrelation mechanism that affects the transport is the time-dependence of the turbulence. If the pressure profile (or temperature profile, in the case of ITG) is non-flat, a diamagnetic drift velocity is present and the frequencies are non-zero (as in eq. \eqref{4}). The problem is that these frequencies are proportional to the wavenumber $k_y$, which has a direct effect on the $\textbf{E}\times \textbf{B}$ component of the motion in the radial direction. In fact, it can be shown that a linear dispersion relation $\omega(\textbf{k}) = k_y V^*$ is equivalent with a turbulent potential drifting along the $y$-direction with velocity $V^*$, exactly as desired. The effect of this drift characteristic of the turbulence is that some of the particles remain trapped, but exhibit elongated trajectories, while others are trapped, but are carried away along the $y$-axis; the latter is equivalent with open trajectories. This break of topology induces a strong suppression of the radial transport, which can make diffusion fall to zero much faster than in the purely frozen case. On the other hand, our frequencies do not follow an exact dispersion relation; this makes them similar to a superposition between a drift motion and a random evolution of the field. The latter is a decorrelation mechanism which tends to increase the transport. 

The main component of the radial transport in eqs. \eqref{1.1} -- \eqref{1.4} is the $\textbf{E} \times \textbf{B}$ drift, that amounts to $\sim \hspace{-3pt} \partial_y\phi$. This implies that a gross measure of the radial velocities experienced by the particles is $v_x\propto \Phi/\lambda_y$. At the level of a quasilinear analysis, the diffusion coefficient can be evaluated as $v_x^2\tau_c$. Depending on which of the decorrelation mechanisms is faster, $\tau_c$ can be estimated as $\lambda_x/v_x$, $\lambda_z/v_{\parallel}$, or even something more complex derived from the frequency dispersion relation.

Now we can understand the behavior of the radial diffusion with each of the five parameters (\Cref{fig_8a,fig_8b,fig_8c,fig_8d,fig_8e}). If the perpendicular decorrelation is the one that dominates, then the Kubo number \cite{palade_pom_ghit_scaling} is $K_\star = \Phi\lambda_x/\lambda_y$. It is known that the turbulent transport is usually anomalous and $\dif^\infty\sim K^\gamma$. This explains why the asymptotic radial diffusion drops roughly as $\lambda_y^{-1}$ (Fig. \ref{fig_8c}) and grows almost linearly with $\Phi$ and $\lambda_x$ (\Cref{fig_8a,fig_8e}). 

Things are more complicated for the two remaining dependencies, $\lambda_z$ and $V^*$. The decorrelation time is always a complicated mix between all physical processes involved. Although it seems that $\tau_c = \lambda_x/v_x$ gives a good gross estimation of this time, it is not the only contribution. As already discussed, the parallel motion through the turbulent field also decorrelates the trajectories, with a decorrelation time of $\sim \lambda_z/\langle v_{\parallel} \rangle$. In principle, $\dif(t \to \infty) = \dif_{\mathrm{frozen}}(\tau_c)$; this explains the influence of $\lambda_z$ on the diffusion (Fig. \ref{fig_8d}), as it seems to follow the running diffusion profile.

Regarding the dependence with the diamagnetic drift velocity $V^*$, the competition between the two mechanisms is obvious: when $V^*$ is small, its main effect is to decorrelate the trajectories and thus to lower the decorrealtion time and increase the diffusion; when $V^*$ is too large, the particle drift along with the field becomes stronger and the diffusion is lowered (Fig. \ref{fig_8a}).

\section{Conclusions and outlook}
\label{s4}

In the present work, we have investigated NNs as a tool for predicting turbulent transport in a simplified model of tokamak plasmas. The training and validation sets have been obtained through high-resolution test-particle simulations of ion dynamics in turbulent electric fields. The input of the NN is the 5-dimensional set of free parameters of the transport model, $\left\{ V^*, \lambda_x, \lambda_y, \lambda_z, \Phi \right\}$, and the output is the single value of the asymptotic radial diffusion, $\dif^{\infty}$.

The final NN is constructed as an ensemble of 12 individual NNs with different architectures, all optimized with the ADAM method and using a normal Xavier initialization of the weights and biases. The \texttt{tanh} activation function is employed in all nets since, prior to training, the data is rescaled to be constrained between $\pm 1$. 

During the learning phase, the convergence properties of the NN proved to be fast, and the validation predictions, accurate -- both indicators of efficient learning. In the testing phase, the NN predictions were in good agreement with the real data, with an average error of below $2\%$. Moreover, the NN was able to accurately reproduce dependencies with lower degrees of freedom, with one through four of the five parameters fixed, with an overall error of below $2\%$, as well. 

Comparing the NN with a spline interpolation of the TD has shown that the former makes better predictions for both the validation set, and the testing sets with lower degrees of freedom, $V_{1-4}$. Moreover, it is able to accurately extrapolate, with average errors of $\sim \hspace{-3pt} 5\%$. Another key advantage is that NNs can use unstructured data for training, unlike interpolation, which works well when the data is distributed on an equidistant grid; this facilitates adding new data to the initial TD, after it has been created.

Based on these results, we consider that training NNs for predicting turbulent transport in tokamak plasmas proves to be a precise and viable method, both accuracy- and time-wise. This approach shows advantages over other methods, such as interpolation, and is up to $10^6$ times faster than a single test-particle simulation. While the transport model is simplified and the database is limited, the purpose of this article as a proof-of-concept for this method has been reached, establishing the framework for subsequent research in which a more robust model of tokamak plasma dynamics will be employed, and a more adaptable and extensive training database will be built.

\section*{Acknowledgement}

This work has been carried out within the framework of the EUROfusion Consortium, funded by the European Union via the Euratom Research and Training Programme (Grant Agreement No 101052200 — EUROfusion). Views and opinions expressed are however those of the author(s) only and do not necessarily reflect those of the European Union or the European Commission. Neither the European Union nor the European Commission can be held responsible for them.

\bibliographystyle{unsrt}
\bibliography{biblio}

\begin{thebibliography}{10}

\bibitem{gyro_1}
Tobias Goerler, Xavier Lapillonne, Stephan Brunner, Tilman Dannert, Frank
  Jenko, Florian Merz, and Daniel Told.
\newblock The global version of the gyrokinetic turbulence code gene.
\newblock {\em Journal of Computational Physics}, 230(18):7053--7071, 2011.

\bibitem{gyro_2}
AG~Peeters, Yann Camenen, F~James Casson, WA~Hornsby, AP~Snodin, D~Strintzi,
  and Gabor Szepesi.
\newblock The nonlinear gyro-kinetic flux tube code gkw.
\newblock {\em Computer physics communications}, 180(12):2650--2672, 2009.

\bibitem{gyro_3}
Cole~Darin Stephens, Xavier Garbet, Jonathan Citrin, Clarisse Bourdelle,
  K~Lucas van~de Plassche, and Frank Jenko.
\newblock Quasilinear gyrokinetic theory: a derivation of qualikiz.
\newblock {\em Journal of Plasma Physics}, 87(4):905870409, 2021.

\bibitem{palade_fast}
D.~I. Palade and M.~Vlad.
\newblock Fast generation of gaussian random fields for direct numerical
  simulations of stochastic transport.
\newblock {\em Statistics and Computing}, 31(5):60, Aug 2021.

\bibitem{palade_w}
Dragos~Iustin Palade, Madalina Vlad, and Florin Spineanu.
\newblock Turbulent transport of the w ions in tokamak plasmas: properties
  derived from a test particle approach.
\newblock {\em Nuclear Fusion}, 61(11):116031, 2021.

\bibitem{real_time_modeling_1}
J.~Morrow-Jones, M.~Firestone, S.~Jardin, C.~Kessel, and T.K. Mau.
\newblock Use of tokamak dynamics models for digital filtering and control.
\newblock In {\em 15th IEEE/NPSS Symposium. Fusion Engineering}, volume~1,
  pages 219--222 vol.1, 1993.

\bibitem{palade_peaking}
DI~Palade.
\newblock Peaking and hollowness of low-z impurity profiles: an interplay
  between itg and tem induced turbulent transport.
\newblock {\em Nuclear Fusion}, 63(4):046007, 2023.

\bibitem{madi_hidden}
Madalina Vlad, Dragos~Iustin Palade, and Florin Spineanu.
\newblock Hidden lagrangian coherence and memory effects in the statistics of
  hamiltonian motions.
\newblock {\em arXiv preprint arXiv:2306.07639}, 2023.

\bibitem{palade_fast_ions}
DI~Palade.
\newblock Turbulent transport of fast ions in tokamak plasmas in the presence
  of resonant magnetic perturbations.
\newblock {\em Physics of Plasmas}, 28(2), 2021.

\bibitem{larmor_avg_1}
T~Hauff and F~Jenko.
\newblock Turbulent e$\times$ b advection of charged test particles with large
  gyroradii.
\newblock {\em Physics of Plasmas}, 13(10), 2006.

\bibitem{larmor_avg_2}
JM~Dewhurst, B~Hnat, and RO~Dendy.
\newblock Finite larmor radius effects on test particle transport in drift
  wave-zonal flow turbulence.
\newblock {\em Plasma Physics and Controlled Fusion}, 52(2):025004, 2010.

\bibitem{palade_alpha}
A~Croitoru, DI~Palade, M~Vlad, and F~Spineanu.
\newblock Turbulent transport of alpha particles in tokamak plasmas.
\newblock {\em Nuclear Fusion}, 57(3):036019, 2017.

\bibitem{ITG_spectrum_exp_1}
RJ~Fonck, G~Cosby, RD~Durst, SF~Paul, N~Bretz, S~Scott, E~Synakowski, and
  G~Taylor.
\newblock Long-wavelength density turbulence in the tftr tokamak.
\newblock {\em Physical review letters}, 70(24):3736, 1993.

\bibitem{ITG_spectrum_exp_2}
T~Hauff and F~Jenko.
\newblock E$\times$ b advection of trace ions in tokamak microturbulence.
\newblock {\em Physics of Plasmas}, 14(9), 2007.

\bibitem{ITG_spectrum_exp_3}
F~Merz and F~Jenko.
\newblock Nonlinear interplay of tem and itg turbulence and its effect on
  transport.
\newblock {\em Nuclear Fusion}, 50(5):054005, 2010.

\bibitem{ITG_spectrum_exp_4}
Jonathan Citrin, H.~Arnichand, J.~Bernardo, C.~Bourdelle, Xavier Garbet,
  F.~Jenko, Sebastien Hacquin, M.J. Pueschel, and R.~Sabot.
\newblock Comparison between measured and predicted turbulence frequency
  spectra in itg and tem regimes.
\newblock {\em Plasma Physics and Controlled Fusion}, 59, 05 2017.

\bibitem{ITG_spectrum_analytical}
M~Vlad and F~Spineanu.
\newblock Electron heat transport regimes in multi-scale turbulence.
\newblock {\em Physics of Plasmas}, 22(11), 2015.

\bibitem{palade_parallel}
Madalina Vlad, Dragos~Iustin Palade, and Florin Spineanu.
\newblock Effects of the parallel acceleration on heavy impurity transport in
  turbulent tokamak plasmas.
\newblock {\em Plasma Physics and Controlled Fusion}, 63(3):035007, 2021.

\bibitem{palade_pom_2022}
D.I. Palade and L.~Pomârjanschi.
\newblock Effects of intermittency via non-gaussianity on turbulent transport
  in magnetized plasmas.
\newblock {\em Journal of Plasma Physics}, 88(2):905880202, 2022.

\bibitem{palade_pom_ghit_scaling}
D~I Palade, L~M Pomârjanschi, and M~Ghiţă.
\newblock Scaling laws of two-dimensional incompressible turbulent transport.
\newblock {\em Physica Scripta}, 99(1):015201, dec 2023.

\bibitem{ANN_basics}
Jinming Zou, Yi~Han, and Sung-Sau So.
\newblock Overview of artificial neural networks.
\newblock {\em Artificial neural networks: methods and applications}, pages
  14--22, 2009.

\bibitem{ANN_pattern_recognition}
Oludare~Isaac Abiodun, Aman Jantan, Abiodun~Esther Omolara, Kemi~Victoria Dada,
  Abubakar~Malah Umar, Okafor~Uchenwa Linus, Humaira Arshad, Abdullahi~Aminu
  Kazaure, Usman Gana, and Muhammad~Ubale Kiru.
\newblock Comprehensive review of artificial neural network applications to
  pattern recognition.
\newblock {\em IEEE access}, 7:158820--158846, 2019.

\bibitem{ANN_clustering}
Farhat Roohi.
\newblock Artificial neural network approach to clustering.
\newblock {\em Int. J. Eng. Sci.(IJES)}, 2(3):33--38, 2013.

\bibitem{ANN_function_approximation}
Sibo Yang, TO~Ting, Ka~Lok Man, and Sheng-Uei Guan.
\newblock Investigation of neural networks for function approximation.
\newblock {\em Procedia Computer Science}, 17:586--594, 2013.

\bibitem{palade_ANN}
DI~Palade.
\newblock Predicting the turbulent transport of cosmic rays via neural
  networks.
\newblock {\em arXiv preprint arXiv:2307.06062\textup{, 2023; Accepted for
  publication in ``Journal of Cosmology and Astroparticle Physics"}}.

\bibitem{ANN_prediction}
Bihao~H Guo, Dalong~L Chen, Biao Shen, Cristina Rea, Robert~S Granetz, Long
  Zeng, Wenhui~H Hu, Jinping~P Qian, Youwen~W Sun, and Bingjia~J Xiao.
\newblock Disruption prediction on east tokamak using a deep learning
  algorithm.
\newblock {\em Plasma Physics and Controlled Fusion}, 63(11):115007, 2021.

\bibitem{ANN_dynamic_control}
Jonas Degrave, Federico Felici, Jonas Buchli, Michael Neunert, Brendan Tracey,
  Francesco Carpanese, Timo Ewalds, Roland Hafner, Abbas Abdolmaleki, Diego
  de~Las~Casas, et~al.
\newblock Magnetic control of tokamak plasmas through deep reinforcement
  learning.
\newblock {\em Nature}, 602(7897):414--419, 2022.

\bibitem{xavier}
Xavier Glorot and Yoshua Bengio.
\newblock Understanding the difficulty of training deep feedforward neural
  networks.
\newblock In {\em Proceedings of the thirteenth international conference on
  artificial intelligence and statistics}, pages 249--256. JMLR Workshop and
  Conference Proceedings, 2010.

\bibitem{SGD}
Shun-ichi Amari.
\newblock Backpropagation and stochastic gradient descent method.
\newblock {\em Neurocomputing}, 5(4-5):185--196, 1993.

\bibitem{ADAM}
Diederik~P Kingma and Jimmy Ba.
\newblock Adam: A method for stochastic optimization.
\newblock {\em arXiv preprint arXiv:1412.6980}, 2014.

\bibitem{AUG_parameters_1}
R.M. McDermott, C.~Angioni, M.~Cavedon, A.~Kappatou, R.~Dux, R.~Fischer,
  P.~Manas, and the ASDEX Upgrade~Team.
\newblock Validation of low-z impurity transport theory using boron
  perturbation experiments at {ASDEX} upgrade.
\newblock {\em Nuclear Fusion}, 62(2):026006, dec 2021.

\bibitem{AUG_parameters_2}
H~Zohm, J~Adamek, C~Angioni, G~Antar, CV~Atanasiu, M~Balden, W~Becker,
  K~Behler, K~Behringer, A~Bergmann, et~al.
\newblock Overview of asdex upgrade results.
\newblock {\em Nuclear fusion}, 49(10):104009, 2009.

\bibitem{ensemble_ann}
Christopher~M Bishop.
\newblock {\em Neural networks for pattern recognition}, chapter 9.6.
\newblock Oxford university press, 1995.

\bibitem{mathematica}
Wolfram~Research{,} Inc.
\newblock Mathematica, {V}ersion 13.3.
\newblock Champaign, IL, 2023.

\end{thebibliography}

\end{document}